\documentclass[a4paper,10pt]{article}
\usepackage[margin=0.9in]{geometry}
\pdfoutput=1
\usepackage[hidelinks]{hyperref}
\usepackage{fontenc}
\usepackage{algpseudocode}
\usepackage[dvipsnames]{xcolor}
\usepackage{amsmath}
\usepackage{amsfonts,amssymb,amsthm, bbm,braket,mathtools}
\usepackage{mathtools}
\usepackage{xcolor}
\usepackage{graphicx}   % need for figures
\usepackage{amsbsy} % for bold greek letters
\usepackage[numbers]{natbib}
\usepackage{silence}
\usepackage{comment}
\usepackage{algorithm}
\usepackage{soul}
\usepackage{tikz}
\usepackage{todonotes}
\usepackage{cancel}
\usepackage{silence}
\usepackage{bbm}

\newtheorem{theorem}{Theorem}
\newtheorem*{theorem*}{Theorem}
\newtheorem{problem}[theorem]{Problem}

\newtheorem*{lemma*}{Lemma}

\usepackage{caption}
\usepackage{subcaption}
\usepackage{graphicx}

\usepackage{tikz}
\usetikzlibrary{quantikz}
\usepackage{braket}
\DeclareFontFamily{U}{wncy}{}
\DeclareFontShape{U}{wncy}{m}{n}{<->wncyr10}{}
\DeclareSymbolFont{mcy}{U}{wncy}{m}{n}
\DeclareMathSymbol{\Sh}{\mathord}{mcy}{"58} 
\newcommand{\M}{225}
\newcommand{\N}{225}
\newcommand{\T}{112}
\newcommand{\bval}{0.9}
\newcommand{\iter}{5000}

\title{Quantum Phase Estimation without Controlled Unitaries}

\author{Laura Clinton\footnote{laura@phasecraft.io}$\:\:^{1}$, Toby S. Cubitt$^{1,2}$, Raul Garcia-Patron$^{1,3}$, Ashley Montanaro$^{1,4}$,\\ Stasja Stanisic$^{1}$ and Maarten Stroeks\footnote{m.e.h.m.stroeks@tudelft.nl}$\:\:^{1,5}$
\\ \\ \large{$^1$Phasecraft Ltd.}\\ \large{$^2$University College London}\\ \large{$^3$University of Edinburgh}\\ \large{$^4$University of Bristol}\\ \large{$^5$Delft University of Technology}}
\date{}

\begin{document}

\maketitle

\begin{abstract}
In this work we demonstrate the use of adapted classical phase retrieval algorithms to perform control-free quantum phase estimation. We eliminate the costly controlled time evolution and Hadamard test commonly required to access the complex time-series needed to reconstruct the spectrum. This significant reduction of the number of coherent controlled-operations lowers the circuit depth and considerably simplifies the implementation of statistical quantum phase estimation in near-term devices. This seemingly impossible task can be achieved by extending the problem that one wishes to solve to one with a larger set of input signals while exploiting natural constraints on the signal and/or the spectrum. We leverage well-established algorithms that are widely used in the context of classical signal processing, demonstrating two complementary methods to do this, vectorial phase retrieval and two-dimensional phase retrieval. We numerically investigate the feasibility of both approaches for estimating the spectrum of the Fermi-Hubbard model and discuss their resilience to inherent statistical noise.
\end{abstract}

\section{Introduction}
\label{sec:introduction}

The prediction of spectral properties is a key aspect in the discovery of novel materials and advancing our understanding of chemical properties and reactions \cite{duckett2000foundations}. The evaluation of spectral properties of large many-body systems is a computationally intensive task, limited by numerical bottlenecks associated with classical simulation of time-evolution \cite{PhysRevX.5.041041}.
Quantum computers have the potential to make such simulations tractable, offering the opportunity to revolutionize  material science and chemistry \cite{Bauer2020}. Algorithms based on quantum phase estimation offer a route towards obtaining such spectral properties, albeit requiring one to perform many costly time-evolution simulation steps \textit{controlled} on one or several ancillary qubit(s), making it difficult to run on current \textit{fault-prone} quantum computers. Therefore, eliminating or reducing the need for controlled time-evolution in quantum phase estimation has attracted a lot of attention, as it could render the prediction of spectral properties accessible to near-term devices. 

A significant step in this direction was the proposal of statistical phase estimation (see e.g. \cite{SommeEarly, Somma_2019, O’Brien_2019}) which reduces quantum phase-estimation to the computation of a time series $f(t)=\bra{\psi}e^{iHt}\ket{\psi}$ on a quantum device followed by classical post-processing to obtain an estimate of the spectrum of $H$ with similar resolution as standard quantum phase estimation for most Hamiltonians. As shown in Figure 1.a, statistical phase estimation reduces the required quantum circuit to two variants of a Hadamard test circuit, giving the real and imaginary parts of $f(t)$. This result has spurred a plethora of theoretical work \cite{Lin2022,Wan2022,Wang2023}, followed by experimental demonstrations \cite{Riverlane2023, Google21}, which can be adapted to estimation of more general spectral quantities such as linear response of quantum systems \cite{Baroni2022}, operator-resolved density of states \cite{Lukin2023} or Green functions \cite{Troyer2016}, to give some examples. Even if statistical phase-estimation represents indisputable progress toward making quantum phase estimation more reachable for near-term devices, a single-qubit control time-evolution remains a highly non-trivial operation to implement on near-term devices. This makes it often impractical to implement statistical phase estimation on currently existing hardware (beyond trivial-size problems). Its difficulty arises from the significant numbers of qubits and non-trivial circuit depth time-evolution operations that need to be controlled by one (or few) qubit(s), making it extremely sensitive to decoherence, increasing the cost of each time-evolution step, and also significantly increasing the circuit depth of implementing all of the required controlled-time-evolution operations.

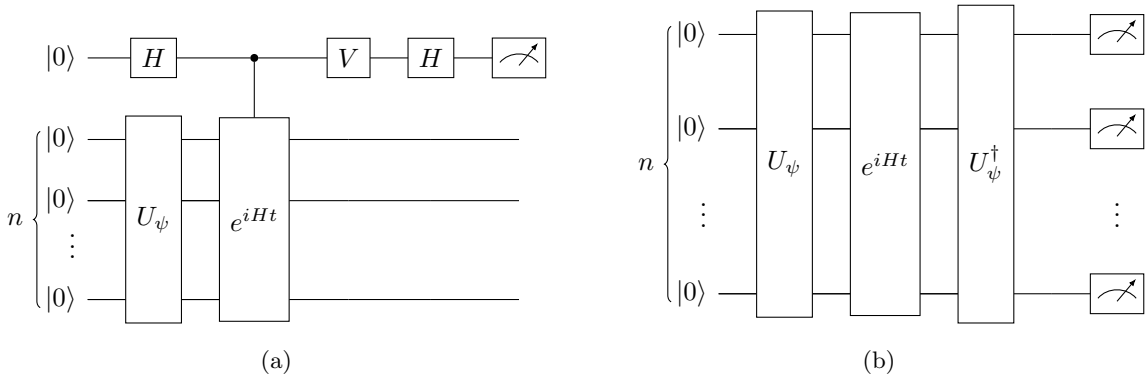
\begin{figure}[t] 
     \centering
     \hspace{-1.0cm}
     \begin{subfigure}[b]{0.5\linewidth}
         \centering
         \begin{quantikz}[thin lines]
            & \lstick{$\ket{0}$}& \gate{H} & \ctrl{1} & \gate{V} & \gate{H} & \meter{} \\
            \lstick[wires=4]{$n$} & \lstick{$\ket{0}$} & \gate[4, nwires=3]{U_\psi} & \gate[4, nwires=3]{e^{iHt}} & \qw & \qw & \qw \\
            & \lstick{$\ket{0}$} & \qw & \qw & \qw & \qw & \qw \\
            & \lstick{\vdots} &  &  &  &  &  \\
            & \lstick{$\ket{0}$} & \qw & \qw & \qw & \qw & \qw 
         \end{quantikz}
         \caption{}
         \label{fig:withancilla}
     \end{subfigure}
     %\hfill
     \hspace{0.3cm}
     \begin{subfigure}[b]{0.4\linewidth}
         \centering
         \begin{quantikz}[thin lines]
            \lstick[wires=4]{$n$} & \lstick{$\ket{0}$} & \gate[4, nwires=3]{U_\psi} & \gate[4, nwires=3]{e^{iHt}} & \gate[4, nwires=3]{U^{\dagger}_\psi} & \qw & \meter{} \\
            & \lstick{$\ket{0}$} & \qw & \qw & \qw & \qw & \meter{} \\
            & \lstick{\vdots} &  &  &  &  & \vdots \\
            & \lstick{$\ket{0}$} & \qw & \qw & \qw & \qw & \meter{}
         \end{quantikz}
         \caption{}
         \label{fig:withoutancilla}
     \end{subfigure}
    \caption{(a) Single-ancillary-qubit quantum phase estimation computing a time series $f(t)=\bra{\psi}e^{iHt}\ket{\psi}$ for time t. The circuit $U_\psi$ prepares the input state $\ket{\psi}$, which is followed by a circuit implementing $e^{iHt}$ controlled by the single-ancillary-qubit. The gate $V$ is an identity ($S^ \dagger$ gate) when estimating the real (imaginary) part of the time series. (b) Ancillary-qubit-free circuit that allows one to estimate $|f(t)|^2 = |\bra{\psi}e^{iHt}\ket{\psi}|^2$.}
    \label{fig:ancillaandancillafree}
\end{figure}

The significant limitation imposed by the \textit{controlled} time-evolution operation has led to an effort to develop algorithmic techniques to retrieve spectra using control-free time evolution operations. Note that the quantity $|\bra{\psi}e^{iHt}\ket{\psi}|^2$ can be obtained with the circuit shown in Figure 1.b, which does not make use of a controlled operation. Unfortunately, this only gives us access to the absolute values of the time-series $f(t)$, while its phases are key in recovering the spectrum.  One way of exploiting the access to $|\bra{\psi}e^{iHt}\ket{\psi}|^2$, is to compute its Fourier transform to retrieve a new spectrum composed of differences between energy levels, rather than obtaining the energies directly, as done in \cite{chan2024algorithmic}. Alternatively, in \cite{Yang2024} the authors use a complex-time evolution, and in \cite{Russo2021} use the knowledge of one eigenvector and eigenenergies, to obtain information about the spectrum using control-free operations. As we detail in the next subsection, each technique has its own limitations and sometimes strong constraints, leaving room for significant improvements.

Retrieving a spectrum while not having (complete) access to the phases of its signal is a well-known problem in classical signal-processing called phase-retrieval \cite{1963Walther}, which has been extensively studied and offers a plethora of different solutions, each one having its own benefits and set of most appropriate applications. The time-series obtained on a digital quantum computer or quantum simulator can be understood as a classical signal and retrieving the spectrum of the many-body problem can therefore benefit from phase-retrieval ideas and algorithms.

In this work we show how the adaptation to quantum problems of two different techniques, \textit{vectorial phase-retrieval} \cite{VPR} and the \textit{two-dimensional phase-retrieval} \cite{fienup82}, allows us to remove the intricate controlled time-evolution currently limiting the scalability of phase-estimation algorithms. Vectorial phase-retrieval works by including measurements of absolute values of additional time series, and -- crucially -- measurements of the absolute values of interferences of the target time series and these additional time series. In addition, we show how to recast the problem as a two-dimensional one, which can be solved with the well known hybrid input-output algorithm \cite{fienup82}. We numerically investigate the feasibility of both approaches for estimating the spectrum of the Fermi-Hubbard model and discuss the resilience of the algorithms to inherent statistical noise.

\subsection{Related work}
In 2020, Russo and collaborators proposed a technique \cite{Russo2021} to measure the energy difference between two eigenvectors of the Hamiltonian that allows one to replace the controlled time-evolution by a simpler time evolution at the cost of having to prepare equal superpositions of both eigenvectors, which for many problems of interest may be intractable. Last year, there were two works that removed the constraint on the input state requirements, making ancillary-free quantum phase-estimation closer to the range of applicability of traditional quantum phase-estimation. In \cite{chan2024algorithmic}, Chan et al.\ leverage the power of the recently proposed classical shadows technique to estimate energy gaps via the computation of expectation values of certain operators as a function of time. A downside of the technique is that it cannot provide the spectrum of $\bra{\psi}e^{iHt}\ket{\psi}$, but rather the spectrum of $|\bra{\psi}e^{iHt}\ket{\psi}|^2$, which leaves the challenge of recovering the eigenvalues of the Hamiltonian from the set of all pairwise energy gaps. Almost at the same time, Yang et al. showed that one can remove the ancillary qubit by combining real-time and imaginary-time evolution \cite{Yang2024}, exploring ideas from complex function analysis. Its main limitation is the implementation of imaginary-time evolution on a quantum device itself, which currently works only for input states with finite correlation length and short evolution times, but improvement in this direction could render the technique more widely applicable. 
Finally, in the recent work \cite{Lukin2023} the authors propose a hybrid classical-quantum algorithm for calculating quantities of interest in many-body spectroscopy that exploits ideas inspired by classical shadows that do not require a controlled time-evolution due to particular properties of the spectral objects of interest. Nonetheless, the approach requires the application of a (potentially shallow) controlled unitary gate independent of the time evolution. In addition, despite the promise of delegating part of the computation to a classical computer, the constraints it imposes in terms of having access to families of states that can be classically emulated may limit its use for arbitrary Hamiltonians.

\subsection{Comparison with controlled time evolution}

There are two key advantages of using phase-retrieval compared with the standard statistical phase-estimation approach that uses controlled time-evolution: (1) It removes a significant source of errors resulting from the sensitivity of the original scheme to the decoherence of the single qubit controlling the dynamics; (2) It requires a smaller number of gates and less circuit depth to implement time evolution compared to its controlled counterpart. Here we summarize in Table \ref{tab:complexities} the reduction in circuit complexity achieved by phase retrieval for three simple examples that capture the essence of the discussion: i) the 1D Transverse-Field Ising Model (TFIM) implemented on hardware with all-to-all connectivity; ii) 1D TFIM with matching hardware connectivity; iii) the $2$D Fermi-Hubbard model with square-lattice hardware connectivity. In each case we calculate the complexity of 1D vectorial phase retrieval. We do not take into account the cost of preparing or uncomputing the initial state $\ket{\psi}$, which we assume to be simple, nor the cost of preparing a superposition of $\ket{\psi}$ and another state\footnote{In our experiments, the superposition used was essentially a GHZ state, but we do not believe this is a fundamental requirement.}. More details on the discussion and derivation can be found in Appendix \ref{sec:comparison}.

In the case of quantum hardware with all-to-all connectivity, a quantum circuit computing a unitary operator $U$ can be converted into a quantum circuit for computing controlled-$U$ by replacing every CNOT gate with a Toffoli gate (requiring 6 CNOT gates each), and every single-qubit unitary $V$ with a controlled-$V$ operator (requiring 2 CNOT gates each). Despite being only a constant factor gain, this can make a significant difference to the ability of near-term quantum computers to obtain meaningful results, by increasing the number of accessible Trotter steps.

\begin{table}[t]
\begin{center}
    \begin{tabular}{|c|c|c|c|c|c|}
        \hline Model & H/w & CNOTs  (PR) & CNOTs (no PR) & Depth (PR) & Depth (no PR) \\
        \hline 1D TFIM & All-to-all & $(2n-2)k$ & $(6n-4)k$ & $4k$ & $2\lceil \log_2 n \rceil + 10k$\\
        1D TFIM & 1D & $(2n-2)k$ & $6(n-1) + (6n-4)k$ & $4k$ & $6\lceil n/2 \rceil + 10k$\\
       % 2d TFIM & 2d & & & &\\
        2D FH & 2D & $32 k n (n-1)/2  $ & $48 k n (n-1)/2   +6 \lceil (n-1)^2 / 2 \rceil$& $32 k$ & $48 k+3 (n-2)$\\
        \hline 
    \end{tabular}
\end{center}
\caption{The cost of implementing $k$ Trotter layers for several Hamiltonians, an $n$ qubit transverse field Ising model and an $n\times n$ spinless Fermi-Hubbard model. ``Depth'' is CNOT depth. Note that the algorithms achieving the minimal CNOT depth and CNOT count may be different. ``2D FH'' is Fermi-Hubbard without spin, where results are approximate. In each case ``PR'' refers to the 1D vectorial phase retrieval approach, in which we do not consider the cost of preparing the initial state.}
\label{tab:complexities}
\end{table}

However, the difference in complexity between the two situations can be significantly greater if there are restrictions on hardware connectivity, as in 1D and 2D architectures.
It is instructive to compare the bounds obtained in a realistic regime for near-term quantum algorithms. As an example, assume that we have a quantum computer with 100 qubits that can execute quantum circuits of CNOT depth at most 100 accurately. Then, based on the algorithmic complexities in Table \ref{tab:complexities}, we could not even implement 1 Trotter step for the 1D TFIM under 1D connectivity if the standard method is used, but we could implement 25 Trotter steps using phase retrieval. The reductions in gate count are less significant than depth, but are still relatively substantial (e.g.\ approximately a factor of 3 for the 1D TFIM).

A key aspect of making phase-retrieval work is to enlarge the problem that we wish to solve in a ``cheap" way, either by enlarging the set of input states included in the procedure, or by implementing additional (comparatively simple) time dynamics. The saving in terms of circuit depth therefore comes at a price of a larger number of circuits to be implemented and required shots.

\subsection{Structure of paper}
Section \ref{sec:summary} summarizes our results on the adaptation of both vectorial phase-retrieval and two-dimensional phase-retrieval to statistical quantum phase-estimation with a final comparison of the performance of both techniques.
This is followed by a conclusion and discussion (Section \ref{sec:conc-disc}).
The manuscript also contains three Appendices on the technical preliminaries (Appendix \ref{sec:prelims}),
and the full details of our adaptation of vectorial phase-retrieval (Appendix \ref{sec:VPR}) and two-dimensional phase-retrieval (Appendix \ref{sec:2DPR}) to statistical phase-estimation.

\section{Summary of results}
\label{sec:summary}

The classical post-processing in statistical phase-estimation is a standard signal processing problem:  One is provided with a noisy version of some complex-valued signal $f(t)$, which is typically measured at integer multiples of some time increment $\Delta t$ (resulting in a discrete time-series\footnote{We use square brackets to stress that the function takes discrete input.}$f[t]$). The goal is to approximately determine the Fourier transform  $F(\omega)$ of the continuous signal, typically corresponding to some quantity of interest such as the spectrum of a Hamiltonian. This is a non-trivial task due to noise and to finite measurement time $T$. There exists extensive literature on this problem with applications to multiple areas of engineering, such as optical imaging, spectroscopy and audio processing \cite{oppenheim1997signals,crystallography,astro,optics}. We note that in the statistical phase-estimation scenario, the largest measurement time $T$ is bounded by the largest circuit depth we can afford in the simulation of  the \textit{controlled} time evolution unitary $e^{iHT}$.

The goal of phase-retrieval as presented in classical signal processing literature is to reconstruct the signal from the absolute values of its discrete Fourier transform. At first sight, this is an impossible task, as the phases are key to achieve the reconstruction. As with any no-go result, the question is which rules one needs to change to circumvent it. This can be successfully achieved by extending the problem that one wishes to solve to one with a larger set of input signals and exploiting natural constraints on the signal and/or the spectrum. 
A key non-trivial insight is to understand when this generalized and constrained problem has a unique solution that recovers all phases of the signals. If such an insight can be used to develop an algorithm that solves the problem in the ideal noiseless case, a non-trivial aspect is to design algorithms that work efficiently 
in realistic regimes of noise and have reasonable computational cost. This work is devoted to designing these algorithms in the scenario of ancillary-free quantum phase-estimation (QPE). 
It is important to note that phase-retrieval literature defines (the time-series) \textit{unique solution} up to unavoidable trivial ambiguities corresponding to a multiplication by a global phase, a time-shift, and time-reversal complex-conjugation of the signal $f[t]$. The last two generate the same spectrum, while the global phase 
on the time-series induce a shift of the spectrum that has no real physical impact, as experimentally we only have access to energy differences.\footnote{There is an additional subtlety due to the fact that phases are defined over a bound domain and a shift becomes a cyclic permutation. A proper map of the energies to phases can ensure
that the smallest value of the spectrum remains above 0 and the smallest below $2\pi$, resulting in a potentially shifted but not distorted spectrum.}

The adaptation of phase-retrieval algorithms to the quantum-phase-estimation scenario is unfortunately not a trivial ``plug \& play". First it requires a conceptual adaptation of the framework. The standard
problem in classical phase-retrieval literature consists of reconstructing the phases $e^{i \theta[t]}$ of $f[t]$ while having access to $|F [\omega]|$, where in our case we know $|f[t]|$ and want to recover $e^{i \theta [t]}$ (and thereby $F[\omega]$) instead. 
Because of the symmetries of the standard classical signal processing problem where both signal and its spectrum are potentially complex functions, we can most of the time circumvent this slight difference by just exchanging the role of time and frequency in most phase-retrieval algorithms. 
Secondly, one needs to find an enlarged set of time-series, by enlarging the set input states or inducing additional dynamics, that combined with natural constraints on the spectrum, such as its non-negativity or the fact that it has a well-defined support, guarantee an efficient recovery of the phases $e^{i\theta[t]}$ (we shall come back to this point below and further details are provided in Sections \ref{sec:VPR} and \ref{sec:2DPR}).
This will finally allow approximate reconstruction of our initial spectrum of interest, but -- importantly -- also of the phases of the time series $f[t]$. In this manuscript, we show how to re-design two well-known phase-retrieval techniques, vectorial phase-retrieval and two-dimensional phase-retrieval, to address the problem of ancillary-free quantum phase-estimation. Vectorial phase-retrieval \cite{VPR} achieves its goal by including measurements of the absolute values of multiple well-behaved time-series, and -- importantly -- the time-series of their interferences. Where standard vectorial phase-retrieval makes use of two signals (the target signal and one additional signal) and their interferences, we generalize to multiple additional signals %and adapt the convex-optimization relaxation 
to improve the algorithm's resilience to noise in the quantum phase-estimation framework. Secondly, we show how to transform a one-dimensional time series (corresponding to time evolution of an input state under a Hamiltonian) into a two dimensional phase-retrieval problem by adding a dummy Hamiltonian $H_D$ that commutes with $H$ and produces non-trivial dynamics on the input state of interest. 
We then exploit the well-known \textit{Hybrid Input Output} algorithm to retrieve the target spectra.

It is important to point out that our objective is never to recover the individual eigenvalues in the spectrum of a given Hamiltonian, which is not practical even for a fault-tolerant computer due to the exponential increase of the number of spectral peaks with the size of the system. A fair goal for this work is to compare the phase-retrieval approaches to what could be achieved with a noiseless statistical phase-estimation algorithm that has access to controlled time evolution but with the same realistic constraint on the time interval over which the simulation can be implemented. 

We stress that the goal of phase-retrieval is solely to reconstruct the phases of the time series and we could use different metrics to quantify the quality of its recovery. Because our ultimate aim is to use the time-series to later generate a quantum phase-estimation spectrum related to a given many-body problem, we will use the recovery of the spectrum as a good metric of performance of the phase-retrieval problem. In order to keep the analysis simple we just implement a recovery relying on a DFT of the time-series, which being a unitary transform will preserve the distance between the ideal time-series and phase-retrieved counterpart. However, we note that the discrete Fourier transform is not necessarily the most accurate reconstruction of the spectrum of a Hamiltonian. Typically, classical post-processing techniques like the one developed in \cite{Somma_2019} can be used to obtain a more useful description of the spectrum.

In the rest of the document we will use dimensionless time and frequency variables, i.e., we will use the notation $f[j]$, where $j\in\mathbb{N}$ and $t_j=j \Delta t$, and $F[k]$, where $k\in\mathbb{N}$ and $\omega_k = 2 \pi k/T$, and we have defined $\Delta t = T/N$. We will be interested in instances of the Fermi-Hubbard model over a line or square lattice graph $G$, where $V$ is the set of vertices and $E$ the edges, where each lattice site, i.e., vertex, has two spin modes $\sigma \in \{\uparrow,\downarrow\}$.  The Hamiltonian reads 
\[ H = -\tau \sum_{\braket{i,j}\in E,\sigma} \left( a^\dag_{i\sigma} a_{j\sigma} + a_{j\sigma}^\dag a_{i\sigma} \right)+ U \sum_{v\in V} n_{v\uparrow} n_{v\downarrow}, \]
where the first term of the Hamiltonian consists of hopping terms among modes of same spin while the second sum corresponds to interactions between particles of opposite spin at the same lattice site. 
Throughout this work we will take $\tau=1$ and $U=4$, which is an intermediate coupling regime where the model exhibits non-trivial behaviour \cite{FH3}. When referring to states, we are -- throughout this work -- referring to qubit states, which are related to the fermionic states by a Jordan-Wigner transformation. For later reference, we note that a standard basis qubit state corresponds to a position-spin-basis Slater determinant.

\subsection{Vectorial phase retrieval}
\label{sec:summaryVPR}

Vectorial phase retrieval \cite{VPR} is a one-dimensional phase retrieval technique. One-dimensional phase retrieval techniques aim to solve a problem that, clearly, is ambiguous in general. Namely, without any further constraints, \textit{every} assignment of phases, denoted by $\{x_{j}\}_{j=0}^{N-1}$, constitutes a time series $\{|f[j]|x_{j}\}_{j=0}^{N-1}$ that is consistent with the absolute value measurements $\{|f[j]|\}_{j=0}^{N-1}$.
Therefore, in order to (approximately) solve the one-dimensional phase retrieval problem, one has to include additional constraints or measurements. In the vectorial phase retrieval framework, this is done by including absolute value measurements of interferences between the target time series $f$ (which we shall denote by $f_1$ in the remainder of this section) and another (secondary) time series $f_2$. 
To be more precise, vectorial phase-retrieval resolves the ambiguity in the retrieval of the phases of the signal $f_1$ by measuring not only its absolute values $|f_1[j]|$, but also those of a secondary signal $|f_2[j]|$ and their interferences $|f_1[j]+f_2[j]|$ and $|f_1[j]+if_2[j]|$. The vectorial phase retrieval problem has a \textit{unique} solution -- up to trivial ambiguities -- in the noiseless scenario, provided that (1) $f_1$ and $f_2$ are \textit{spectrally independent} (i.e., their $z$-transform have no common roots in the complex plane), and (2) the discrete Fourier transforms of $f_{1}$ and $f_{2}$ (denoted resp. by $F_{1}$ and $F_{2}$) have no support outside some finite interval (of size at most $N/2$ or, equivalently, of size at most $\pi$ in terms of frequency) \cite{VPR}. 

In the setting of control-free quantum phase estimation, we take $f_1[j] = \bra{\Phi}\exp(ij H\Delta t)\ket{\Phi}$ and $f_2[j] = \bra{\Phi}\exp(ij H\Delta t)\ket{\psi}$ (for some state $\ket{\psi}$ different from $\ket{\Phi}$), so that $f_1[j] + f_2[j] = \bra{\Phi}\exp(ij H\Delta t)\big(\ket{\Phi} + \ket{\psi} \big)$ and $f_1[j] + if_2[j] = \bra{\Phi}\exp(ij H\Delta t)\big(\ket{\Phi} + i\ket{\psi} \big)$. The constraint of $F_1$ and $F_2$ only being supported on a frequency interval of size $\pi$ can be satisfied by an appropriate choice of $\Delta t$. Note that for standard statistical phase estimation routines, such a constraint should also be satisfied, except that there the support should be of size $2\pi$.

Remark that in our scenario we are interested in retrieving the phases of the time series $f_{1}$, which can then -- for instance -- be used to obtain $F_1$. 
Two important factors that lead to one only obtaining an approximation of the correct assignment of phases $\{x_{j}\}_{j=0}^{N-1}$ in a practical setting are (1) the noise on the learned signals $|f_{1}|$ and $|f_{2}|$ and (2) the fact that $F_{1}$ and $F_{2}$ fail to have their support restricted to a well-defined finite interval. 
This second factor is caused by spectral leakage of $F_{1}$ and $F_{2}$ away from the eigenvalues of $H$, which may spread outside the expected restricted support. This can be particularly aggravated when the state $\ket{\Phi}$ and distribution of supported eigenvalues might be such that $F_{1}$ and $F_{2}$ decay slowly near either end of the $[0,\pi]$ interval.

We find that to ensure that the approximate solution obtained through vectorial phase retrieval is accurate, one has to ensure that the scenario is such that $F_{1}$ and $F_{2}$ have \textit{approximately} well-defined support. This can generally be done by for instance increasing the total evolution time (at the expense of deeper circuits implementing the time evolution) or making informed choices for the reference states or for $\ket{\Phi}$ (when the problem we aim to solve allows for some freedom in the choice of the input state). Our numerical investigations suggest that vectorial phase retrieval performs relatively well even when $F_1$ and $F_2$ do not have well-defined support.
 
In what follows we summarize our implementation of the vectorial phase retrieval technique (and its application to control-free quantum phase estimation), more details and derivations can be found in Section \ref{sec:VPR}.

\subsubsection{Vectorial phase retrieval as an optimization problem}
\label{sec:summaryvprrelaxation}

The key aspect behind the vectorial phase-retrieval technique is the realization that the phases of
$f_1[j]$ and $f_2[j]$ are related by the relation $e^{\theta_1[j]} = e^{\theta_2[j]}\: G[j]$, where
\begin{equation}
     G[j]:= \frac{|f_1[j]+f_2[j]|^{2} + i|f_1[j]+if_2[j]|^{2} - (1+i)\big( |f_1[j]|^{2} + |f_2[j]|^{2} \big)}{2\:|f_1[j]|\:|f_2[j]|},
\end{equation}
is fully accessible from the measurement data. This allows us to define an optimization problem 
over the variable $\mathbf{y}\in \mathbb{C}^{2N}$ with unit absolute value entries, encoding the estimates of the phases $\{e^{\theta_1[j]},e^{\theta_2[j]}\}$. The cost function to be optimized consists of two contributions. The first contribution quantifies how close we are from satisfying the relative phase constraints $e^{\theta_1[j]} = e^{\theta_2[j]}\: G[j]$ and reads,
\begin{equation}\label{eq:Cost-int}
    Q_{\text{interference}}(\mathbf{y}) := \sum_{j=0}^{N-1} \:\bigl\lvert\:y_{j} - y_{N+j}\:G[j]\:\bigr\rvert^{2}. 
\end{equation}
As will be argued in Section \ref{sec:VPR} (and in \cite{VPR}), imposing these phase constraints is not sufficient for (approximate) recovery of the phases in general. Therefore, the cost function also contains a second contribution. This additional contribution quantifies how close we are from $F_1$ and $F_2$ being zero outside the support $\{0,1,\ldots,\sigma\}$ (for some integer $\sigma \in \{0,1,\ldots,\lfloor N/2\rfloor-1\}$, the \textit{exact} value of which is unknown in a general setting). In an ideal noiseless scenario, this contribution is minimized if $F_1$ and $F_2$ are zero on the domain $\{\sigma+1,\sigma+2,\ldots,N-1\}$. In reality, one of the factors making this constraint not fully satisfied is the inevitable spectral leakage of the discrete Fourier transform, that will lead to $F_{1}$ and $F_{2}$ having support outside of the supported frequency range.
Nonetheless, as we will demonstrate, the approach still recovers the phases relatively accurately provided that $|F_1[k]|$ and $|F_2[k]|$ decay quickly for $k > \sigma$. 
%\begin{equation}
%    Q^{(s)}(\mathbf{y}) := Q^{(s)}_{\text{support}}(\mathbf{y}) + Q_{\text{interference}}(\mathbf{y}),
%\end{equation}
To impose that the solution (approximately) has the correct support size, we define a family of support cost functions, one for each $s\in\{0,1,\ldots,N-1\}$ (because, again, the true support size $\sigma$ might be unknown, or not even well-defined), defined as 
\begin{equation}\label{eq:Cost-support}
    Q^{(s)}_{\text{support}}(\mathbf{y}) =\: \sum_{k=s}^{N-1} \Bigl\lvert \sum_{j=0}^{N-1}|f_1[j]|\: y_{j}\exp\big[-i2\pi jk/N\big]\Bigr\rvert^{2} + \sum_{k=s}^{N-1} \Bigl\lvert \sum_{j=0}^{N-1}|f_2[j]|\: y_{N+j}\exp\big[-i2\pi jk/N\big]\Bigr\rvert^{2},
\end{equation}
quantifying how far the spectrum is from being zero on the region $\{s,s+1,\ldots,N-1\}$. 

As we explain in more detail in 
section \ref{sec:VPR}, one can define a family of total cost functions, combining the support component and the interference component, which can be expressed as a non-negative quadratic form
\begin{equation}
    Q^{(s)}(\mathbf{y}):= Q^{(s)}_{\text{support}}(\mathbf{y}) + Q_{\text{interference}}(\mathbf{y}) = \mathbf{y}^{\dagger}A_{s}^{\dagger}A_{s}\:\mathbf{y},
\end{equation}
where $A_{s}\in \mathbb{C}^{(2(N-s)+N)\times 2N}$ entries are detailed in Section \ref{sec:VPR}. Remark that $Q^{(s)}(\mathbf{y})$ is equal to zero for all $s\in\{\sigma+1,\sigma+2,\ldots,N-1\}$ if $\mathbf{y}$ is the unique solution of the ideal noiseless vectorial phase-retrieval problem (i.e., for which $|F_1[k]|$ and $|F_2[k]|$ are indeed equal to zero for $k>\sigma$). When $|F_1[k]|$ and $|F_2[k]|$ are not exactly equal to zero for $k>\sigma$, the minimum does not equal zero for any $s \in \{0,1,\ldots,N-1\}$. Typically, however, it will decay quickly as a function of $s$ around the optimal choice of $s$. 
As presented in Figure \ref{fig:vecPRsummary}, finding the solution to the optimization problems at different values of $s$ typically allows one to obtain an estimate of an approximate $\sigma$, which in turn can be used to approximately retrieve the phases of $f_1$.

In any case, the optimal assignment of phases $\mathbf{y}$ for a given $s$ can be found by minimizing the function $Q^{(s)}(\mathbf{y})$ over all assignments of \textit{phases} $\mathbf{y}$, which is not a convex optimization problem due to the constraints on the entries of $\mathbf{y}$ (i.e., that each entry has unit absolute value). To approximate the optimal solution in a practical setting, one will have to relax the non-convex optimization over phase assignments. The relaxation that we will employ is the problem of optimizing $Q^{(s)}(\mathbf{y})$ over all vectors $\mathbf{y}\in \mathbb{C}^{2N}$ for which $||\mathbf{y}||_{2} = \sqrt{2N}$, instead of optimization over $\mathbf{y}$ whose entries have unit absolute value.\footnote{Remark that the relaxation we use is slightly different to the one implemented in \cite{VPR}.} 
%Note that the optimal solution satisfies $||\mathbf{y}||_{2}=\sqrt{2N}$, mapping 
Hence, we relax the problem to the problem determining the smallest eigenvalue of a matrix.

Empirical investigation, illustrated in Figure \ref{fig:vecPRsummary}, has shown us that in order to make the phase-retrieval more resilient to noise, it is convenient to generalize the problem by including $R>1$ \textit{secondary} signals and their respective interferences with the target signal (as opposed to including just one such secondary signal). This leads to an equivalent expression for the cost function, now expressed in terms of $\tilde{A}_{s}\in \mathbb{C}^{((R+1)(N-s)+RN)\times (R+1)N}$ (where $\tilde{\cdot}$ is used to stress that noisy input data is used in the definition of $\tilde{A}_{s}$) and $\mathbf{y}_{R}\in \mathbb{C}^{(R+1)}$. Employing an equivalent relaxation as the one just discussed, we optimize over all vectors $\mathbf{y}\in \mathbb{C}^{(R+1)N}$ for which $||\mathbf{y}||_{2}=\sqrt{(R+1)N}$, corresponding to the following eigenvalue problem.
\begin{equation}
    \hspace{-0.4cm} \min_{\substack{\mathbf{y}\in\mathbb{C}^{2N}\text{ s.t. } ||\mathbf{y}||_{2} = \sqrt{(R+1)N}}} \hspace{-0.2cm}\tilde{Q}^{(s)}(\mathbf{y}) = (R+1)N\:\lambda_{\min}\big( \tilde{A}_{s}^{\dagger}\tilde{A}_{s} \big).
\label{eq:relaxationnoisysummary}
\end{equation}
Our numerical investigations suggest that for relatively small noise magnitudes, the vector $\mathbf{y}$ obtained by solving Eq. \eqref{eq:relaxationnoisysummary} will generally be close to the ideal solution. We do note that the entries of the optimal vector $\mathbf{y}$ are not necessarily phases (i.e., have unit absolute value), nonetheless, we approximate $\{f_1[j]\}_{j=0}^{N-1}$ by $\{|f_1[j]|y_{j}\}_{j=0}^{N-1}$ because our numerical investigations suggest that this generally gives a more accurate reconstruction compared to the solution that is obtained when rounding to a vector whose entries are phases. For a more detailed discussion of this latter fact, we refer to Section \ref{sec:vecpr_numerics}.

\subsubsection{Implementation}

To demonstrate the technique, we apply the vectorial phase retrieval framework to the state evolution signals generated by a $(1\times 5)$ spinful Fermi-Hubbard Hamiltonian with $|U/\tau| = 4$. The target initial state is taken to be a (standard-basis) state at half filling, where we map to quibts under a Jordan-Wigner transformation. We consider a scenario in which the number of secondary states is one (i.e., $R=1$), and a scenario in which we include ten secondary states (i.e., $R=10$). The \textit{secondary} states are created as follows. Take the target state and select at most $p<n/2$ (distinct) pairs of qubits that are unequally occupied, then flip the states on these qubits. We obtain the time series at $N = 300$ points in time (with $N_{\text{samples}} = 10^6$ shots per point in time), with a time increment $\Delta t = 0.133$.

Note that superpositions of the target state and a given secondary state can be produced using similar circuit to those generating a multi-qubit GHZ state followed by local $X$ gates, which requires $O(\log n)$ layers of two-qubit gates (given all-to-all connectivity). This is similar to a method that can be used to reduce the cost of controlled time-evolution on hardware with restricted connectivity, or the depth on hardware with all-to-all connectivity (see Table \ref{tab:complexities}). While one should take this additional cost into account when comparing complexities, we believe that it is not a fundamental limitation of the method, as different secondary states could be used.

\begin{figure}[h]
        \subfloat[$N_{\text{samples}} = 10^{6}$.]{%
            \includegraphics[width=.596\linewidth]{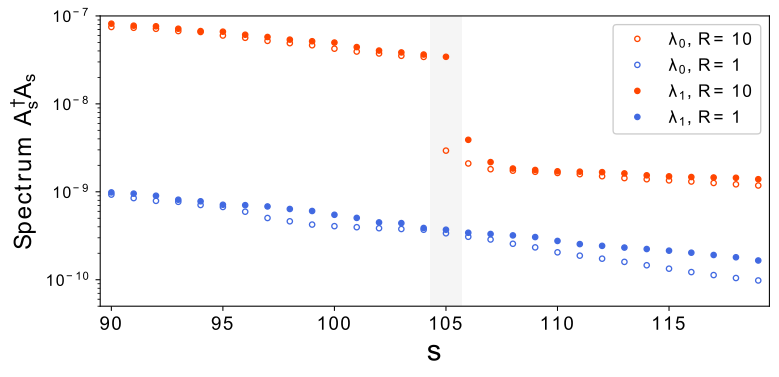}%
            \label{fig:multRc}%
        }\hfill
        \subfloat[$N_{\text{samples}} = 10^{6}$, $s=105$.]{%
            \vspace{-0.05cm}
            \includegraphics[width=.404\linewidth]{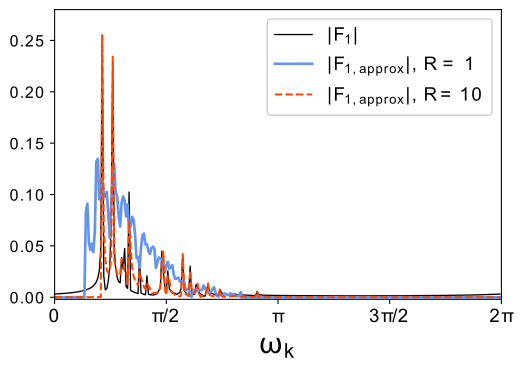}%
            \label{fig:multRd}%
        }
        \caption{(a) The smallest two eigenvalues of $A_{s}^{\dagger}A_{s}$ for $R=1$ (blue) and $R=10$ (red) in the noisy setting corresponding to $N_{\text{samples}} = 10^{6}$ as a function of $s$; (b) Comparison between the reconstructed spectrum for $R=1$ (blue) and $R=10$ (red) versus the exact solution (black) (obtained by evaluating the smallest eigenvector of $A_{s}^{\dagger}A_{s}$ at $s = 105$). Note that in (a) it can be seen that, in the $R=10$ scenario, the smallest eigenvalue drops rapidly around $s = 105$ while it is the opposite for $R=1$.}
%and the spectral gap is relatively large at $s = 105$.}
        \label{fig:vecPRsummary}
\end{figure}

Figure \ref{fig:vecPRsummary}a depicts the smallest two eigenvalues of the $\tilde{A}_{s}^{\dagger}\tilde{A}_{s}$ matrix, whose smallest eigenvector is used as an estimate for the assignment of phases of the time series entries in both scenarios. 
Our investigations suggest that:
\begin{enumerate}
    \item The two smallest eigenvalues are a proxy for the quality of the retrieval of phases. That is, the smaller the smallest eigenvalue, the smaller the value of the two residuals Eq. (\ref{eq:Cost-int}) and (\ref{eq:Cost-support}) above. 
    \item A large gap opening between the two smallest eigenvalues suggests a larger overlap between the smallest eigenvector and the true assignment of phases. This being motivated by the fact that in the ideal noiseless case the solution is unique. 
\end{enumerate}

Typically, the smallest eigenvalue of $\tilde{A}_{s}^{\dagger}\tilde{A}_{s}$ drops rapidly as a function of $s$ at a given value $s^{*}$ (at $s = 105$ in Figure \ref{fig:vecPRsummary}a), which our numerical investigations suggests is an indicator of an accurate retrieval of phases around that value of $s$. Note that both features mentioned above
are present in the $R=10$ scenario, leading to a good reconstruction, but not in the $R=1$ scenario, which does not provide an accurate reconstruction, as one can see in Figure \ref{fig:vecPRsummary}b (showing the associated estimate of the spectrum $|F_1|$ at $s = 105$).

More generally, our investigations suggest that the noise resilience of the vectorial phase retrieval algorithm is improved by taking $R>1$. We note that increasing $R$ increases the amount of classical data to be processed, the running time of the classical post-processing algorithm and the number of quantum circuit shots required. The trade-off between resources cost and quality of solution will therefore play a role in the final choice of $R$. In addition, it is important to cleverly pick the secondary quantum states in a general setting, so that the circuit generating them and their superposition with the target state is a shallow circuit, so that it does not significantly increase the depth of the quantum circuit needed for the implementation.

\subsection{Two-dimensional phase-retrieval}

It is well-known in the phase-retrieval literature that two dimensional problems are more resilient to non-trivial ambiguities of the solution \cite{Bruck,kogan}.
The non-trivial ambiguities in the $1$D problem arise from the fact that the $z$-transform of the signal is a univariate reducible polynomial.
In contrast, reducible multivariate polynomials have measure zero, which implies that the phase retrieval problem in the multi-dimensional case typically has a unique solution, unless disguising a single variable polynomial as a multivariable one.
Nonetheless,  this result strongly suggests that two-dimensional phase-retrieval should work in practice, an intuition confirmed by a broad range of applications in optical imaging and image processing that demonstrate its practical value \cite{Bendory2017,Osherovich2012}.

We embed our problem -- which is naturally one dimensional -- into a larger $2$D problem so that we can exploit the benefits of two-dimensional phase-retrieval.
The full details of this method can be found in Section \ref{sec:2DPR}, but in summary we introduce an additional dummy Hamiltonian $H_D$ and a second independent time variable $z$.
%\st{A new discretized time series can then be defined }
%\begin{equation}
	%f[j,l] \coloneqq \bra{\psi}e^{i t_j H  }e^{i z_l H_D   }\ket{\psi},
	%\label{eq:2d-sig}
%\end{equation} 
A new signal $f[j,l]$ can be defined from samples of
\begin{equation}
	f(t,z) = \bra{\psi}e^{i t H  }e^{i z H_D   }\ket{\psi},
	\label{eq:2d-sig}
\end{equation} 
at the discrete set of times $t_j =j \Delta t$ and $z_l =l \Delta z$,  where $\Delta t = T/N$ and $\Delta z = T'/M$, where for simplicity we choose  $T'=T$ \footnote{Note that $T=T'$ is not a fundamental requirement and other choices are possible and potentially beneficial.}. This lets us define a matrix $f[j,l]$ whose entries can roughly be thought of as samples of this function at these discrete times. See Section \ref{sec:2DPR}
for the precise definition of the discrete signal $f[j,l]$, as it also includes a windowing function and a specific ordering of these elements.

\begin{figure}[t]
	\begin{subfigure}{.45\linewidth}
		\centering
		\begin{algorithmic}[1]
			\Procedure {Hybrid}{$|f|$, $\beta$, $L$, $F^1$}
			
			\For {i = 1, 2, \ldots L}
			\State $f^i = \mathcal{DFT}^{-1} (F^i) = |f^i| e^{i \text{arg}(f^i)}$
			\State $\tilde{f}^i = |f| e^{i \text{arg}(f^i)}$
			\State $\tilde{F}^i = \text{real}(\mathcal{DFT} (\tilde{f}^i))$
			\If{$\tilde{F}^i[k,m] \leq 0$}
			\State $F^{i+1}[k,m]=F^{i}[k,m] - \beta \tilde{F}^{i}[k,m]$
			\Else
			\State $F^{i+1}[k,m]=\tilde{F}^i[k,m]$
			\EndIf
			\EndFor
			\EndProcedure
		\end{algorithmic}
	\end{subfigure}
	\hspace{1cm}
	\begin{subfigure}{.45\linewidth}
		\centering
		\begin{tikzpicture}[node distance=2.0cm, auto]
			\node (tl)[draw] {$F^{i}$};
			\node (t) [right of=tl,draw,fill=Apricot] {$\mathcal{DFT}^{-1}$} ;
			\node (tr) [right of=t,draw] {$f^{i}$} ;
			\node (cl) [below of=tl,fill=Apricot,draw] { Real and Positive};
			\node (bl) [below of=cl,draw] {$\tilde{F}^{i}$};
			\node (b) [right of =bl,fill=Apricot,draw]{$\mathcal{DFT}$};
			\node (br) [right of =b,draw]{$\tilde{f}^{i}$};
			\node (cr) [below of = tr,fill=Apricot,draw]{$|f[j,l]|$};
			\draw[->] (tl) to node {} (t);
			\draw[->] (t) to node {}(tr);
			\draw[->] (tr) to node {}(cr);
			\draw[->] (cr) to node {}(br);
			\draw[->] (br) to node {}(b);
			\draw[->] (b) to node {}(bl);
			\draw[->] (bl) to node {}(cl);
			\draw[->] (cl) to node {}(tl);
		\end{tikzpicture}
	\end{subfigure}
	
	\caption{Pseudocode (left) and schematic (right) for Fienup's hybrid input-output algorithm in four steps. The $i-$th iteration starts with a candidate spectrum $F^i$: Step (1) transform it into a time series $f^i$ candidate via an inverse DFT; Step (2) generates a new $\tilde{f}^i$ that has same phase as $f^i$
    and satisfies $|\tilde{f}[j,l]|=|f[j,l]|$; Step (3) transform it back to the Fourier domain; Step (4) take the real part and then
    implement the update rule in the schematic to impose positivity of the spectrum.}\label{fig:HIOsummary}
\end{figure}
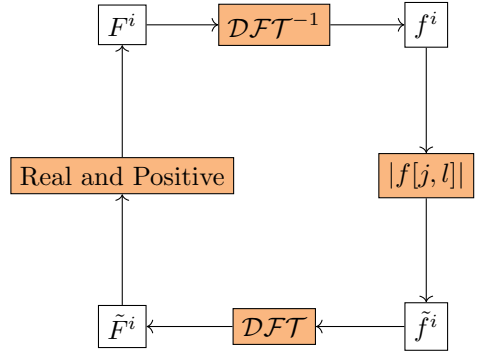
In this work we apply one of the oldest and most widely used phase retrieval algorithms to this new $2$D problem, the hybrid input-output algorithm (HIO) proposed by Fienup \cite{fienup82} in 1982. 
HIO is known to perform well when $F[k,m]$ -- the DFT of $f[j,l]$ -- is real and positive.
This is because this knowledge can be used to drive the HIO algorithm.
These constraints can be imposed on the problem very naturally, as we detail in Section \ref{sec:2DPR}:
\begin{enumerate}
    \item Positivity can be ensured by choosing a $H_D$ which commutes with $H$, and by multiplying the time-series by a filter that has a positive discrete Fourier transform. For simplicity in our implementation we choose a triangular window and multiply by this to form a discrete array $f[j,l]$.
    \item We ensure $F[k,l]$ is real by exploiting the fact that $f[j,l]=f^*[N-j,M-l]$, which will follow from the definition of $f[j,l]$  (see Section \ref{sec:2DPR}).
\end{enumerate}
We can then incorporate this additional knowledge about the solution into the phase retrieval algorithm.
HIO also makes it relatively simple to add additional constraints exploiting symmetries of the problems or partial information we may have on the solution, as we show below for a specific example.
 
As schematized in Figure \ref{fig:HIOsummary},
Fienup's hybrid input-output algorithm (HIO) \cite{fienup82} involves transforming back and forth between the Fourier (left) and object domain (right) interspersed by projections that guarantee the satisfaction of the time-series values $|f[j,l]|$ in the object domain and approximately guarantee the $F[k,m]$ constraints in the Fourier domain. The algorithm is non-linear and of an iterative nature, where the constraint on the spectrum defines a new element $F^{i+1}$ with the update rule shown at the left side of Figure \ref{fig:HIOsummary}, where the parameter $0 \leq \beta \leq 1$ is carefully selected to optimize the performance. 
We will now summarise the details of a specific instance of this set-up for the Fermi-Hubbard model.

\subsubsection{Implementation}

We test this algorithm numerically using a $2$D spin Fermi-Hubbard model, with $U=4$ and $\tau=1$. For the dummy Hamiltonian $H_D$ we use the total particle number operator, which under the Jordan-Wigner encoding is simply a sum of single qubit Pauli-$Z$ operators. We use $2 \times 2$ spin Fermi-Hubbard model and set $N = M = \N$, time $T =\T$ and choose $\ket{\psi}$ to be the uniform super-position over computational basis states. $H$ has been normalized so that its eigenvalues lie between $0$ and $\pi$.
Additionally we incorporate knowledge of the phases of $f[0,l]$ as driving constraints, as they can be computed efficiently classically for our specific input state. These points correspond to evolution under $H_D$ only, which can be performed with local gates in any potential experiment.

\begin{figure}[h]
		\centering
		\begin{subfigure}{0.329\textwidth}
			\includegraphics[width=\textwidth]{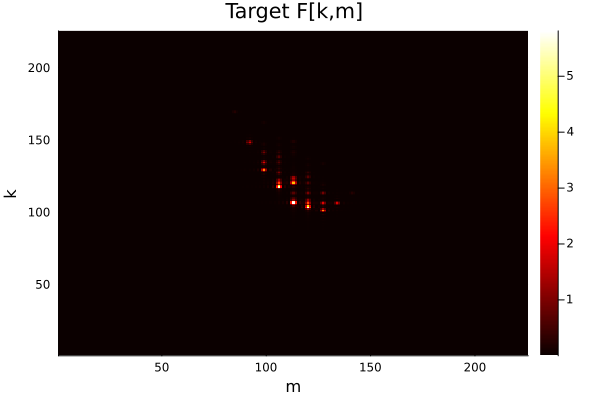}
			\caption{Target spectrum.}
			\label{subfig:target-F}
		\end{subfigure}
		\hfill
		\begin{subfigure}{0.329\textwidth}
			\includegraphics[width=\textwidth]{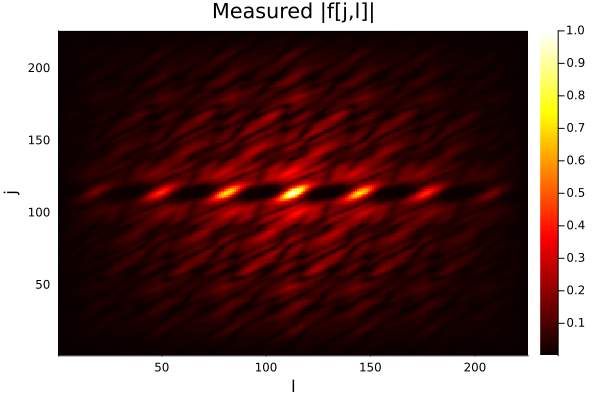}
			\caption{Measured time-series.}
			\label{subfig:meas-f}
		\end{subfigure}
		\hfill
		\begin{subfigure}{0.329\textwidth}
			\includegraphics[width=\textwidth]{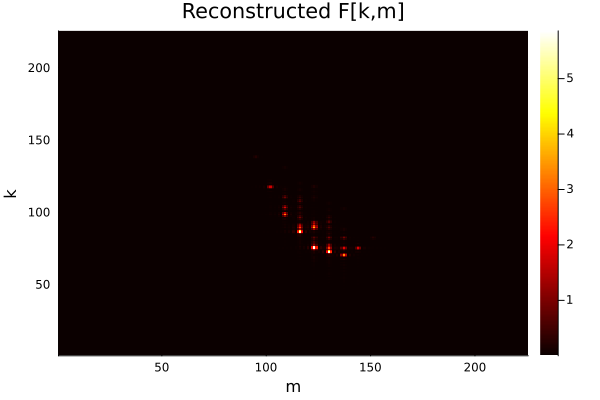}
			\caption{Reconstructed spectrum.}
			\label{subfig:recon-F}
		\end{subfigure}
		\caption{These three plots show an example of $2$D phase retrieval for a $2\times2$ Fermi-Hubbard Hamiltonian with $N \times M = \N \times \M$. Here $T =\T$ and we take $\ket{\psi}$ to be a uniform super-position over computational basis states. $H$ has been normalized so that its eigenvalues lie between $0$ and $\pi$. We have chosen the $H_D$ as the total number operator. This example excludes sampling noise. (\ref{subfig:target-F}) The target $2$D discretized spectrum $F[k,m]$ is shown here as a heatmap; (\ref{subfig:meas-f}) The absolute values $|f[j,l]|$ are shown here, these would be measured using only time-evolution and are the input to the HIO phase-retrieval algorithm; (\ref{subfig:recon-F}) Here we show the result of using HIO on (\ref{subfig:meas-f}), we recover this reconstruction. Observe that the recovered spectrum contains a \emph{trivial} ambiguity, in the form of a shift.}
\label{fig:2d-summary-image}
\end{figure}

The first step is to use the HIO algorithm to reconstruct a $2$D image, summarized in Figure \ref{fig:2d-summary-image}. The left side of Figure \ref{fig:2d-summary-image} shows the target 2D-spectrum $F[k,m]$, which we obtain from a direct numerical calculation of the time-series $f[j,l]$ including its phase.
In the centre we show the absolute values of $|f[j,l]|$, these would be measured using only time-evolution as sketched in Figure \ref{fig:withoutancilla}, and form the input to the HIO algorithm. Observe that we also see the effect of the 2D triangular windowing function, the magnitude of $|f[j,l]|$ decreases as we approach the edge of the plot\footnote{We have plotted the zero frequency in the centre.}.
Finally to the right we show the reconstruction of $F[k,m]$ using the the HIO algorithm with $\beta = \bval$ and $L=\iter$. We can observe a trivial ambiguity in this reconstruction, the final spectrum is translated slightly. Nonetheless the reconstruction is correct up to trivial ambiguities which we correct for when plotting.

\begin{figure} [h]
		\begin{subfigure}{0.45\linewidth}
			\centering
			\includegraphics[width=\linewidth]{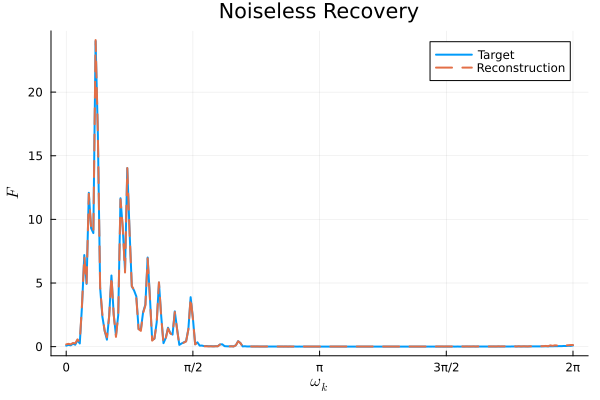}
			\caption{Noiseless}\label{fig:1d-recon}
		\end{subfigure}
		\hfill
		\begin{subfigure}{0.45\linewidth}
			\centering
			\includegraphics[width=\linewidth]{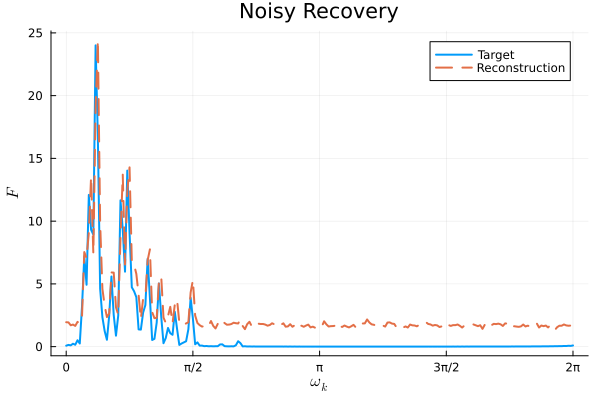}
			\caption{With sample noise.}\label{fig:1d-recon-noise}
		\end{subfigure}
		
		\caption{This figure shows the recovered $1$D spectrum following the image reconstruction in Figure \ref{fig:2d-summary-image}. (\ref{fig:1d-recon}) This shows $F[k]:=\mathcal{DFT}[f[j,0]]$ where $f$ is the inverse discrete Fourier transform of the reconstructed spectrum shown in (\ref{subfig:recon-F}). This is in the noiseless scenario. Note that the ideal solution includes the triangular window by definition. (\ref{fig:1d-recon})  This figure shows the effect of sample noise on this same problem. In this case we use only $1000$ samples per time-series point and still recover the peaks but with a vertical offset in the reconstructed spectrum.}\label{fig:2d-summary}
	\end{figure}

Finally, we recover the $1$D information following this image reconstruction. To do this we apply a DFT to the recovered signal $f[j,0]$, which corresponds to the times-series of $H$ without the dummy Hamiltonian. 
The result of this is shown in Figure \ref{fig:2d-summary}.
We also model the effect of sample noise on this algorithm and find that for a modest number of samples -- $1,000$ per calculation of $|f[j,l]|$ -- the algorithm can still reconstruct the location of the peaks.

\subsection{Comparative Overview of Proposed Methods}

We are now ready to compare the performance of these two approaches in the context of spectral estimation of the Fermi-Hubbard model. We choose a $3 \times 3$ instance of the (spinful) Fermi-Hubbard model (with $|U/\tau| = 4$) and set up the following task. The target spectrum is the ideal solution obtained from the DFT of the ideal time series 
$f[j]=\bra{\psi}e^{ijH\Delta t}\ket{\psi}$ for time step $\Delta t=0.12$ and maximum time evolution of $T=15$, where the time series is obtained using brute-force calculation of $e^{ijH\Delta t}$ (for varying $j$), with the input state being

\begin{align}\label{eq:comparison-state}
        \ket{\psi} = \frac{1}{\sqrt{3}}(\ket{010101010\:010101010} + \ket{110101100\:010100110}+ \ket{011101010\:000110111}),
\end{align}
where the first (last) $9$ bits are associated with the occupation of the spin-up (spin-down) modes.
This state was chosen as it is suitable for our implementations of the vectorial phase retrieval algorithm and the two-dimensional phase retrieval algorithm. It does not provide an unfair advantage to either adaption and -- importantly -- it is not an eigenstate of $H_D$.

To get as close as possible to a fair comparison of both techniques we allocate a budget of $N_S = 5\cdot 10^9$ circuit runs to our simulation of each phase retrieval technique. 
This directly limits the number of circuits runs required for each technique to retrieve each $|f[j]|=|\bra{\psi}e^{ijH\Delta t}\ket{\psi}|$ which translates into shot noise on each value,
which can later affect the performance of the phase-retrieval reconstruction.
More concretely, we model the effect of shot noise induced by the finite number of samples during the statistical estimation of the time-series by sampling from a binomial distribution with success probability $|f[j]|^2$ (See Section \ref{sec:noise}). 

We chose to set up a comparison in this fashion as both methods have a variety of parameters which can be altered to ensure quality performance in the presence of sample noise.
By fixing only the total number of runs we can compare the performance of both methods on the same footing. For example, with vectorial phase-retrieval there is now a trade-off between the number of additional secondary states $R$ permitted and the total number of samples used for obtain an estimate of each $|f[j]|$. The same is true for the two-dimensional phase-retrieval implementation and the parameter $M$. The optimum setting for each implementation remains an open question and depends on the particular problem under consideration. In this work we tried to optimize the choices of parameters for both approaches trying to retrieve a spectrum as close to the target one as possible.

\begin{figure}[h]
        \centering
        
        \begin{subfigure}{0.48\linewidth}
        \centering
        \includegraphics[width=\linewidth]{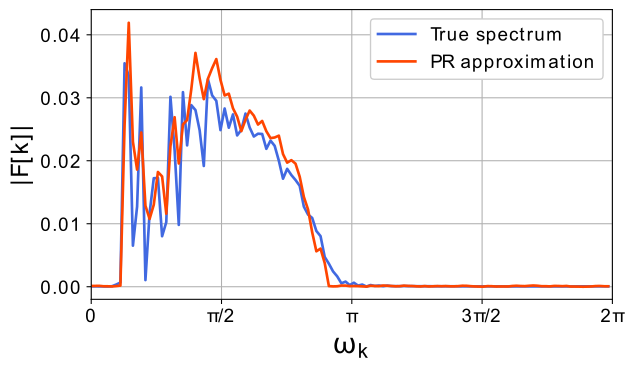}
        \caption{Vectorial phase retrieval.}
        \end{subfigure}
        \begin{subfigure}{0.48\linewidth}
        \centering
        \includegraphics[width=\linewidth]{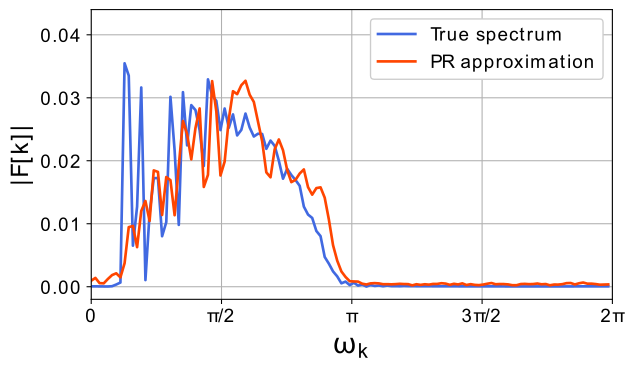}
        \caption{2D HIO.}
        \end{subfigure}
        \caption{This figure compares the two techniques on the same problem with the same sample budget of $5 \times 10^9$. We target the spectrum given the input state $\ket{\psi}$ in Eq. \ref{eq:comparison-state} for the Fermi-Hubbard Hamiltonian $H$ on a lattice of size $3 \times 3$ with on-site interaction strength $U=4$ and hopping strength $\tau = 1$, with time step $\Delta t = 0.12$ and maximum time evolution of $T=15$.
        In both plots the ideal solution is obtained from $f[j]=\bra{\psi}e^ {ijH\Delta t}\ket{\psi}$ and shown in blue; (a) Here we show the estimated spectrum (red) for vectorial phase-retrieval with $R = 40$ and $3.3\cdot 10^5$ shots per point in time $j$ per signal; (b) Here we show the estimate spectrum (red) for two-dimensional phase-retrieval with $M=25$, $\beta = \bval$ and $L=\iter$ and $1.6 \cdot 10^6$ shots per $f[j,l]$.}
        \label{fig:comparaison}
\end{figure}

The phase-recovery is followed by a FFT to reconstruct the Fermi-Hubbard Hamiltonian on a lattice described above, depicted in Figure \ref{fig:comparaison} for comparison. 

For the $2$D implementation we took $M=25$ and used $1.6\times 10^6$ samples per $f[j,l]$.
In conjunction with $\beta=\bval$ and $L=\iter$ iterations the HIO algorithm produces the spectrum shown in Figure \ref{fig:comparaison}. For the vectorial phase retrieval implementation, we used $R = 40$ secondary states, and $6\cdot 10^{5}$ samples per signal per point in time $j$.
We observe that the $2$D HIO method gives a poorer reconstruction in comparison to VPR.
In particular, the HIO algorithm struggles to recover the amplitudes of the peaks.
One can also check that the 1-norm (normalized by $N$) of the vector with entries $|F_{\text{reconstructed}}[k]| - |F_{\text{exact}}[k]|$ is 0.0018 for VPR and 0.0028 for 2DPR,
confirming the intuition from a visual inspection of Figure \ref{fig:comparaison}.
Another observation is that the peaks recovered using $2$D HIO appear slightly broadened
(which generally leads to a lower resolution on the recovered spectrum), due to the windowing function we use to enforce the positivity constraint we use to drive HIO. 

The accurate retrieval of the spectrum using the vectorial phase retrieval algorithm can be attributed to the large number of secondary states employed in the algorithm, leading improved noise resilience (as is outlined in Section \ref{sec:summaryVPR}). Since the vectorial phase retrieval algorithm performs better as the support of the spectrum becomes well-defined, there might be instances of spectra with less well-defined support for which vectorial phase retrieval performs worse. We remark that when HIO is observed to perform well in the noiseless scenario --  such as when the peaks are well defined as in Figure \ref{fig:1d-recon} -- we observe that the corresponding noisy spectrum can typically be reconstructed well with a comparatively low number of samples as in Figure \ref{fig:1d-recon-noise}. We stress that the poorer performance of two-dimensional phase-retrieval in Figure \ref{fig:comparaison} is not necessarily a fundamental fact and one could potentially adapt more advanced modern two-dimensional phase retrieval methods to quantum phase-estimation problem while also exploiting other commuting Hamiltonians $H_D$ that could potentially lead to better results.

The restriction to an $18$-mode Fermi-Hubbard instance is inherent to our numerical modeling demonstration, but we expect that the phase-retrieval algorithms can be applied to system sizes beyond classically tractable ones and can be applied to experimental data from actual quantum hardware. 
The two phase-retrieval techniques are  presented in full detail in Sections \ref{sec:VPR} and \ref{sec:2DPR}.

\section{Conclusion and Discussion}
\label{sec:conc-disc}

We have demonstrated how to design phase-retrieval algorithms for the problem of statistical phase estimation. These algorithms allow for the reconstruction of the phases of the (time evolution) time series of a quantum state $\ket{\psi}$ with respect to a given Hamiltonian $H$, while only having access to their absolute values. To achieve this seemingly impossible goal, one needs to address a more general problem involving (the absolute values of) a larger set of time series, resulting from increasing the set of input states or enlarging the range of time-evolutions available to us. %{\color{ForestGreen}\st{, with the exploitation of constraints on the spectrum that are either physically motivated or easily engineered}}.
To reach our goal we have re-designed two well-known phase retrieval algorithms, vectorial phase retrieval and two-dimensional phase retrieval, which we demonstrate offer good performance in retrieving the spectrum of a Fermi-Hubbard model. 

In the vectorial phase-retrieval scenario, we obtain (absolute values of) time series corresponding to time evolution of multiple input states, but -- more importantly -- time evolution of their superpositions. Finding the right set of quantum states to successfully complement our target input state $\ket{\psi}$ while also being comparatively easy to generate, requires some non-trivial design that will need to be adapted to the Hamiltonian and target input state of interest. We also showed that despite quantum phase estimation being a one-dimensional signal processing problem, one can exploit the benefits of two-dimensional phase-retrieval by adding a quench evolution governed by a dummy Hamiltonian $H_D$ commuting with $H$. 

Despite the two techniques being rather different in design and post-processing algorithms,
there are some analogies in their adaptation to quantum phase estimation.
One could see the time-evolution with dummy Hamiltonian $H_D$ as an alternative way of generating a non-trivial family of time series in addition to the one of the target input state $\ket{\psi}$, in a similar fashion as vectorial phase retrieval exploits multiple (state evolution) time series corresponding to different input states and their superpositions.

We demonstrated that the performance of the algorithms is resilient to shot noise affecting the estimation of the time-series resulting from the limited number of accessible circuit runs.
Further investigation of the impact of circuit imperfection on the performance of the algorithm is left to future work, but as a first proxy one can  model the effect of gate imperfections
by increasing the value of the shot noise.

The adaptation of phase-retrieval to a specific spectrum of a given quantum state $\ket{\psi}$ and Hamiltoian $H$ currently needs a handcrafted approach. It would be interesting to design techniques that can automatize the procedure for arbitrary states and Hamiltonians.

We stress that phase retrieval algorithms recover the phases of the state-evolution time series, of which one learns just the absolute values in the scenario of control-free quantum phase estimation. In some portions of this work, for convenience, we compare the discrete Fourier transform of the retrieved time series to that of the true time series to judge the accuracy of the phase retrieval. However, we note that the discrete Fourier transform is not necessarily the best way to recover the spectrum of a Hamiltonian from its time-series. Typically, classical post-processing techniques like the one developed in \cite{Somma_2019} can be used to obtain a more accurate and useful description of the spectrum. We note that since phase retrieval algorithms recover the phases of the state-evolution time series, their output could be processed with more advanced techniques, such as the post-processing algorithm from \cite{Somma_2019}. 

We note that despite the Hybrid Input-Output algorithm used in two dimensional phase-retrieval working well in-practice, it is highly probable that relaxing the problem to a convex optimization would bring additional benefits, from convergence guarantees, to better running times, scaling and potentially better resilience to noise. We leave this question open for future work.

Finally, we observe that although our phase retrieval algorithms can achieve nontrivial reductions in quantum circuit complexity in many cases, whether this is useful in practice requires consideration of the tradeoff between circuit complexity and number of shots required, since our algorithms generally increase the latter. It is an interesting question whether this increased number of shots required can be significantly reduced. There is a wealth of advanced techniques in the classical phase retrieval literature that could be explored and evaluated to improve the results that we present.

\subsection*{Acknowledgments}

This work was supported by Innovate UK (grant no. 44167). AM acknowledges funding from the European Research Council (ERC) under the European Union's Horizon 2020 research and innovation programme (grant agreement No. 817581).

\section{Appendix A: Technical Preliminaries}\label{sec:prelims}

In this appendix we first introduce basics concepts of statistical phase-estimation and signal processing that will be of particular interest to readers not familiar 
with phase-estimation and its signal-processing material. Secondly, we give details on the Fermi-Hubbard Hamiltonian that we use as a benchmark and few technicalities related to our simulations.

\subsection{Definitions}

In what follows we adopt the following continuous Fourier transform convention
		\begin{align}
			F(\omega)=\mathcal{FT}[f] &:= \int_{-\infty}^{\infty} f(t) e^{-  i\omega t} dt \hspace{1cm}
			f(t)=\mathcal{FT}^{-1}[F] := \frac{1}{2\pi}\int_{-\infty}^{\infty} F(\omega) e^{  i\omega t} d \omega,
		\end{align}
		and the discrete Fourier Transform  definition as
		\begin{align}
			F[k]= \mathcal{DFT}[f] &:= \sum_{j  =0}^{N-1} f[j] e^{-i 2 \pi \frac{ k j}{N} } \hspace{1cm}
			f[j]=\mathcal{DFT}^ {-1}[F]:=\frac{1}{N} \sum_{k=0}^{N-1} F[k] e^{i 2 \pi \frac{ k j}{N}} . 
		\end{align}

For a spectrum $F(\omega)$ to be a real function it needs to satisfy $F(\omega)=F^*(\omega)$ for all $\omega$. It is easy to check from the definitions above that this is only possible if
\begin{equation}
    F(\omega)-F^*(\omega)=\int\left[f(t)-f^*(-t)\right]e^{i\omega t}dt=0,
\end{equation}
which holds if and only if $f(t)=f^*(-t)$.

Let's consider an Hamiltonian $H=\sum_iE_i\Pi_i$, where $E_i$ is the $i$-th eigenvalue of the Hamiltonian and $\Pi_i$ is the projector into its eigenspace. In the case of a rank 1 projector we have $\Pi_i=\ket{E_i}\bra{E_i}$.  Its time evolution reads $e^ {iHt}=\sum_ie^ {iE_i t}\Pi_i$ and Tr$[\rho \Pi_i]=p_i$ is the amplitude of spectrum of $\rho$ at energy $E_i$, where the full spectrum 
of $\rho$ can be written in a compact form as $F(\omega)=2 \pi \sum_i p_i\delta(\omega- E_i)$, where the positivity of $p_i$ implies that of the spectrum $F(\omega)$
%\stas{is there a minus missing in this bracket?}
The Hadamard test in Figure \ref{fig:withancilla} used in statistical phase-estimation allows the computation of the complex amplitude 
\begin{equation}\label{eq:HadamardTest}
    f(t)={\rm Tr}[\rho e^{iHt}],
\end{equation}
for different times $t$, i.e., time series.  It is easy to see that 
$ f(t)={\rm Tr}[\rho e^{iHt}]=f^*(-t)$, due to $\rho$ being hermitian and the cyclic property of the trace, which proves that the spectrum is real. We will now show that it is also positive in the ideal case.

\subsubsection{Positivity of the spectrum}

Applying the Fourier transform we obtain 
\begin{equation*}
    \mathcal{FT}[f(t)]= \int {\rm Tr}[\rho \sum_i e^ {iE_it}\Pi_i] e^{-i\omega t} dt,
\end{equation*}
which using linearity of the trace and some rearrangements leads to 
\begin{equation*}
    \mathcal{FT}[f(t)]= \sum_i  {\rm Tr}[\rho \Pi_i] \int  e^{-i(\omega-E_i) t} dt
\end{equation*}
which using the definition of the amplitudes of the spectrum and $\int e^{-i(\omega-E_i) t} dt= 2 \pi \delta(\omega-E_i)$ leads to
\begin{equation*}
    \mathcal{FT}[f(t)]= 2 \pi \sum_i  p_i\delta(\omega-E_i)=F(\omega).
\end{equation*}

In what follows we assume the Hamiltonian eigenvalues $E_i$ satisfy the condition $E_i\in [0,2\pi]$.
Otherwise, assuming the Hamiltonian maximum and minimum eigenvalue energies, or providing a decent upper and lower bounds, one can always re-scale the Hamiltonian accordingly.

\subsection{Basics concepts on signal processing and Fourier transforms}

\subsubsection{Time-window and resolution}

In practice continuous signals $f(t)$ can only be probed inside a finite time window $[-T/2,T/2]$.
The simplest windowing $\hat{f}(t)=f(t)\Pi_T(t)$ results from the multiplication of the original signal $f(t)$ and a time-window function 
\begin{equation} 
\Pi_T(t) = \begin{cases}
1, & |t|\leq T/2 \\
0, & |t|>T/2
\end{cases}
\end{equation}

Using the convolution theorem, relating the Fourier transform of a product to the convolution of its respective Fourier transforms ($\mathcal{FT}[A\cdot B]=\mathcal{FT}[A]*\mathcal{FT}[B]$), together with the Fourier transforms of the step function 
\begin{equation}
    \mathcal{FT}[\Pi_T(t)](\omega)=\frac{2 \sin \left(\frac{T \omega }{2}\right)}{\omega }  = T\text{sinc}(T\omega/2)
\end{equation}
where we define $\mathcal{FT}[f(t)](\omega)=F(\omega)$, leads to 
\begin{equation}
    \mathcal{FT}[\hat{f}(t)](\omega)= F(\omega)* T \text{sinc}(T\omega/2).
    \label{eq:FTtimewindow}
\end{equation}
We will use the notation $\hat{F}(\omega)$ for the convolution of $=F[\omega]*\text{sinc}(T\omega)$, resulting from the broadening of the spectral peaks due to the finite-size time window.

\paragraph{Resolution}
The convolution $\hat{F}[\omega]=F[\omega]*\text{sinc}(T\omega/2)$ generally leads to a broadening of the spectral peaks of $F(\omega)$. The main lobe width of $\text{sinc}(T\omega/2)$ being $4\pi/T$, it is easy to see that a time window of size $T$ can generally only resolve energy gaps of order $O(1/T)$. Therefore to have full resolution of the spectrum one needs to do time-evolution up to $T$ inverse proportional to the smallest gap between eigenvalues in the spectrum. 

The choice of $T$ gives an upper-bound on how good the quality is of the recovered spectrum, i.e., even under ideal reconstruction conditions one can at best retrieve $=F[\omega]*\text{sinc}(T\omega)$. For Hamiltonians 
with gaps scaling beyond $1/\text{poly}(n)$, it is impossible to reconstruct the exact spectrum and retrieving the coarser version $\hat{F}[\omega]$ will be the only option. 

\paragraph{Alternative time-windows}
Remark that despite the ideal spectrum $F(\omega)$ being positive the convolution with a sinc function, resulting from the finite-size time window, may lead to a spectrum that has some slightly negativities. In our 2D phase-retrieval algorithm this could be an issue for the convergence 
of our algorithm. We will therefore use a different time window that will ensure the target spectrum in our reconstruction being all positive. The simplest way to achieve this is via a triangle time-window, resulting from the convolution of two rectangular time-windows $\Pi_T(t)$, which has a squared sinc function as Fourier transform. Unfortunately, this doubles the width of the main lobe of its Fourier spectrum further widening the recovered spectrum. There is a whole literature on 
optimized filters that would allow to limit the widening of the spectrum while still guaranteeing its positivity, we leave that for future work.

\subsubsection{Sampling and aliasing}
In practice continuous signals $f(t)$ can only be probed inside a finite time window $[-T/2,T/2]$
and for a finite number $N$ of time values. We will assume these samples are evenly spaced with time $\Delta t$.
Note that in this discussion we always choose an odd number of samples $N$ and set the spacing to be $\Delta t=T/N$. 
We have odd time values as $t=0$ is also included and we are probing up to time $t_{N-1} = (N-1)/N \times T/2$. 
The discretised sampling over a finite range is modeled via the multiplication of the initial signal $\hat{f}(t)$ by a Dirac comb 
\begin{equation} \label{eq:Diraccomb}
\Sh_{\Delta t}(t) = \sum_{j\in \mathbb{Z}} \delta(t - j\Delta t)
%D_{\Delta T}(t) = \sum_{n\in \mathbb{Z}}\delta(t-n\Delta T),
\end{equation}
leading to the new discretised and finite signal
\begin{equation}
g(t)=f(t)\Pi_T(t)\Sh_{\Delta t}(t).
\end{equation}
Using the convolution theorem together with the Fourier transforms of the Dirac comb 
\begin{equation}
    \mathcal{FT}[\Sh_{\Delta t}(t)](\omega)=\frac{2\pi}{\Delta t}\Sh_{2\pi/\Delta t}(\omega)=\frac{2\pi}{\Delta t}\sum_{k\in \mathbb{Z}}\delta\left(\omega-k\frac{2\pi }{\Delta t}\right),
\end{equation}

together with eq.(\ref{eq:FTtimewindow}) we obtain \
%\begin{align}
%\mathcal{FT}[g(t)](\omega) 
%    =2\pi\frac{T}{\Delta t}
%    \sum_{k\in \mathbb{Z}} F(\omega)*\text{sinc}(T\omega)*\delta\left(\omega-k\frac{2\pi }{\Delta t}\right).
%\end{align}

\begin{equation}
    G(\omega) \coloneqq \mathcal{FT}[g(t)](\omega)=2\pi\frac{T}{\Delta t}
    \sum_{k\in \mathbb{Z}} F(\omega)*\text{sinc}(T\omega/2)*\delta\left(\omega-k\frac{2\pi }{\Delta t}\right).
\end{equation}

One can simplify the previous equation using the properties of delta functions and their convolutions 
%\begin{equation}
%    \mathcal{FT}[g(t)](\omega)=2\pi N
%    \sum_{k\in \mathbb{Z}} \hat{F}\left(\omega-\frac{k}{\Delta t}\right).
%\end{equation}
and using the definition $\hat{F}(\omega)=F(\omega) * T\text{sinc}(T\omega/2)$ to get
\begin{equation}
    G(\omega)=\frac{2\pi}{\Delta t} 
    \sum_{k\in \mathbb{Z}} \hat{F}\left(\omega-k\frac{2 \pi}{\Delta t}\right).
\end{equation}

\paragraph{Aliasing}
Due to the discrete-time nature of the sampling, the spectrum $\mathcal{FT}[g(t)](\omega)$ becomes periodic function of period $2\pi/\Delta t$. The celebrated Nyquist-Shannon theorem states that sampling at twice the bandwidth of the signal guarantees an unique recovery of the continuous function. If one samples at a lower rate, the weights of higher frequency lines are shifted to the lower part of the spectrum, an effect called aliasing. In an ideal scenario (and unrealistic) scenario of an infinite time-window and the spectrum of $H$ having largest absolute value frequency $\omega_{\text{max}}=\max\{|\omega|\}$, it would be sufficient to guarantee the condition $2 \Delta t \cdot \omega_{\text{max}}<1$ to have perfect reconstruction and zero aliasing. In practice, depending on the choice of time-window and the finite value of $T$ we always observe some amount of aliasing, but a good choice of sampling rate, time-window shape and having access to large enough $T$ will help make aliasing small or even negligible.

\subsubsection{Connecting sampled time-series and discretized spectrum via a DFT}

The algorithms we use exploit discretized version of time-series and spectrum connected via a DFT.
In what follows we review this connection where the discretized spectrum can be understood from a periodic extension of the time-series.

We \emph{choose} as set of discrete frequencies $\omega_k =2 \pi k/T$ where $k= 0, 1, \ldots N-1$ and define a length $N$ vector $F[k] := \mathcal{FT}[g(t)](\omega_k) = G(\omega_k)$, whose entries are samples of the continuous time Fourier transform of $g(t)=f(t)\Pi_T(t)\Sh_{\Delta t}(t)$ at points $\omega_k$.
We begin by writing the Fourier transform of our sampled function as
\begin{align}
        F[k] &\coloneqq  \mathcal{FT}[g(t)](\omega_k) \\
        &= \int_{-\infty}^\infty f(t)\Pi_T(t)\Sh_{\Delta t}(t) e^{- i \omega t}dt ,
\end{align}
where using the definition of the Dirac comb in eq.(\ref{eq:Diraccomb}) and $\int f(t)\delta(x-a)dt=f(a)$
we obtain
\begin{equation}
	F[k]= \sum_{j= -\infty}^{\infty} f(j\Delta t) \cdot \Pi(j \Delta t) e^{-i    \omega_k j \Delta t} 
\end{equation}
which using the discretisation notation $\hat{f}[j]:=f(j \Delta t)\Pi(j \Delta t )$ and the fact that $\omega_k j \Delta t= 2\pi j k/N$ leads to 
\begin{equation}
	F[k] = \sum_{j= -\infty}^{\infty} \hat{f}[j] e^{-2 i \pi  k j/N }.
\end{equation}
The connection between the DFT and samples of the the continuous functions at $\omega_k$ and $t_j= j \Delta t$ is almost complete.
We have obtained the conventional DFT frequencies, however we have an infinite sum in the above expression.

We deal with this as follows.
One can replace the sum over integers by the following double sum
\begin{align}
	F[k] &= \sum_{j' \in \mathbb{Z}} \sum_{j=0}^{N-1} \hat{f}[j- j' N] e^{-i(2\pi  k j/N + 2\pi j' k) } \\
	&=   \sum_{j=0}^{N-1} \left(\sum_{j' \in \mathbb{Z}}  \hat{f}[j- j' N] \right)e^{-i2\pi k j/N  }, \\
	%&= \sum_{n=0}^{N-1} f[n] e^{-i 2\pi  k n/N  },
\end{align}
then using the definition 
\begin{align}
	f[j] :=  \sum_{j' \in \mathbb{Z}} \hat{f}[j - j' N].
\end{align}
leads to
\begin{equation}
	F[k] = \sum_{j=0}^{N-1} f[j] e^{-i 2\pi  k j/N  }.
\end{equation}
This shows that $F[k]$ is actually the DFT of $f[j]$, which is simply an $N$ periodic version of the windowed samples $\hat{f}[j]$.

Notice that $f[j]$ is a periodic sequence where the values within the window are repeated with period $N$. The previous exposition provides an interpretation of the discretized time-series linked to a discrete spectrum via the DFT, in a similar fashion as occurred for their continuous counterparts. Provided we properly choose $T$ and $\Delta t $ to avoid aliasing and limit leakage, we can safely work on a discrete setting using the DFT to move between the sampled time-series and a discretized version of the spectrum.

\subsection{Fermi-Hubbard simulation and technicalities of our numerical experiments}\label{sec:noise}

In this manuscript, we benchmark our phase-retrieval algorithms for performing statistical QPE  on spectrum recovery of the Fermi-Hubbard model. 

\paragraph{Fermi Hubbard} We will be interested in instances of the Fermi-Hubbard model over a square lattice graph $G$, where $V$ is the set of vertices and $E$ the edges, where each lattice site, i.e., vertex, has two spin modes $\sigma \in \{\uparrow,\downarrow\}$.  The Hamiltonian reads 
\[ H = -\tau \sum_{\braket{i,j}\in E,\sigma} \left( a^\dag_{i\sigma} a_{j\sigma} + a_{j\sigma}^\dag a_{i\sigma} \right)+ U \sum_{v\in V} n_{v\uparrow} n_{v\downarrow}, \]
where the first term of the Hamiltonian consists of hopping terms among modes of same spin while the second sum corresponds to interactions between particles of opposite spin at the same lattice site. This model is of great interest, since despite its simple form it demonstrates interesting phenomena found in more complex materials and molecular Hamiltonians \cite{FH1, FH2, FH3}. 
Throughout this work we will take $\tau=1$ and $U=4$, which s an intermediate coupling regime where the model exhibits non-trivial behaviour \cite{FH3}. All simulations in this work are performed on the qubit level, by applying a Jordan-Wigner transformation to the fermionic model.

\paragraph{Time-evolution simulation}
In order to compute time-series of the form $|\bra{\psi}e^{iHt}\ket{\phi}|^2$ (forming the input to our phase retrieval algorithms) and time-series $\bra{\psi}e^ {iHt}\ket{\phi}$ (which we use to compare to our phase-retrieved solution), one needs to simulate the time-evolution of the Fermi-Hubbard model $e^{iHt}$. We will be using numerical simulations to obtain the time-series, by brute-force calculation of $e^{iHt_j}$ with $t_{j} = j \Delta t$. Then, for some states $\ket{\psi}$ and $\ket{\phi}$, we simply evaluate the expressions $f[j] = \bra{\psi} e^{iHt_j} \ket{\phi}$ and $|f[j]|$.  Due to the current limitation on near-term quantum computers, we will be interested to explore states that have shallow preparation circuits, as detailed in Sections \ref{sec:VPR} and \ref{sec:2DPR}.

\paragraph{Sampling Noise}
A crucial aspect of the practical implementation of phase-retrieval algorithms is their resilience to noise. We model the sampling noise on time series $|\bra{\phi}\exp(ij\Delta t H)\ket{\psi}|^{2}$ (for some $\ket{\psi}$ and $\ket{\phi}$) by sampling from a binomial random variable. In an experimental setting, one can estimate $|\bra{0^n}U_{\phi}^\dagger\exp(ij \Delta t H)U_{\psi}\ket{0^n}|^{2}$ (where $U_\psi$ and $U_{\phi}$ are state preparation unitaries s.t. $\ket{\psi}=U_\psi\ket{0}$ and $\ket{\phi}=U_\phi\ket{0}$) by acting with $U_{\psi}^{\dagger}\exp(ij \Delta t H)U_{\psi}$ on $\ket{0^n}$, and simply performing standard-basis projective measurements on each of the $n$ qubits. The estimate is then given by the fraction of the number of times that output $\ket{0}^{\otimes n}$ is obtained and the total number of circuit executions $M$. Therefore, we model the sampling noise by estimating $|\bra{\phi}\exp(ij\Delta t H)\ket{\psi}|^{2}$ by the number of successes $M_{0}$ (divided by $M$) obtained in a binomial random variable realization with success probability $|\bra{\phi}\exp(ij \Delta t H)\ket{\psi}|^{2}$ and $M$ trials.

\paragraph{Trivial Ambiguities}
We call operations that leave the absolute value of the time series unchanged trivial ambiguities. Throughout this paper, we ignore these trivial ambiguities and therefore match the recovered signal to the true time-series $f\in \mathbb{C}^{N}$ up to: (1) a global phase $e^{i\phi}f$; (2) linear phase shifts $\{e^{-i2\pi jm/N}f[j]\}$; (3) complex conjugation $f^{*}$, or combinations thereof. These respectively induce the following operations on the DFT of $f$: (1) multiplication by a global phase; (2) translation by $m$; (2 combined with 3) reflection and complex conjugation. Note that a shift in the energies, provided that the support of the spectrum is bounded, can be ensured to not lead to confusion in the ordering of the recovered peaks. The shape of the spectrum will be correct up to a constant shift.

\section{Appendix B: Vectorial phase retrieval}
\label{sec:VPR}

In this section we give a detailed presentation of vectorial phase retrieval (v-PR) (see \cite{VPR}) and its adaptation to the quantum scenario, which we discussed briefly in Section \ref{sec:summaryVPR}. This problem corresponds to a version of the one-dimensional phase retrieval problem, where one assumes access to particular additional input in the form of interference measurements. 

\subsection{Vectorial phase retrieval with a single interference signal}

Let us first discuss under which circumstances the (single-interference-signal) v-PR problem has a unique solution. This understanding of the uniqueness of the v-PR problem was used in \cite{VPR} to develop a method for obtaining estimates of the phases of the time series $\mathbf{f}\in \mathbb{C}^{N}$.  

In the v-PR framework, the (non-trivial) ambiguity of the one-dimensional phase retrieval problem is reduced by not only obtaining absolute value measurements $\{|f_1[j]|\}$ of a time series $\mathbf{f}_1 \in \mathbb{C}^{N}$, but also measurements $\{|f_{2}(j)|\}$ of some other time series $\mathbf{f}_{2} \in \mathbb{C}^{N}$, and -- crucially -- of the absolute values of the time series corresponding to sums of $\mathbf{f}_{1}$ and $\mathbf{f}_{2}$. It was shown in \cite{BeinertPhaseRetrieval} that under some relatively mild conditions on the time series $\mathbf{f}_1$ and $\mathbf{f}_2$, and assuming \textit{exact} access to the aforementioned absolute value measurements, the resulting phase retrieval problem has a unique solution. The problem of \textit{finding} this unique solution efficiently was addressed in \cite{VPR}. 

Let us define the vectorial phase retrieval problem more exactly. Afterwards, we will discuss under which circumstances the problem has a unique solution, and how one can in practice obtain an accurate estimate of the full time series using the vectorial phase retrieval algorithm \cite{VPR}. 
\begin{problem}
    \normalfont{[Vectorial phase retrieval (v-PR)]} Given measurements of the absolute values 
    \begin{equation}
        |f_1[j]|^2,\:|f_{2}[j]|^2,\: |f_{3}[j]|^2:=|f_1[j]+f_{2}[j]|^2,\: |f_{4}[j]|^2:=|f_1[j]+if_{2}[j]|^2,
    \end{equation}
    at times $j=0,1,\ldots,N-1$, determine 
    \begin{equation}
        F_1[k] := \sum_{j=0}^{N-1}f_1[j]\exp\big[-i2\pi jk/N\big], \: F_{2}[k] := \sum_{j=0}^{N-1}f_{2}[j]\exp\big[-i2\pi jk/N\big],
    \end{equation}
    at $k=0,1,\ldots,N-1$.
\label{prob:simpleVPR}
\end{problem}
\noindent
In the setting of control-free quantum phase estimation, the input signals take the following form. 
\vspace{-0.75cm}
\begin{center}
\begin{align}
    f_{1}[j] =&\: \bra{\Phi}\exp(i\Delta tHj)\ket{\Phi}, \nonumber \\
    f_{2}[j] =&\: \bra{\Phi}\exp(i\Delta tHj)\ket{\psi}, \nonumber \\
    f_{3}[j] =&\: \bra{\Phi}\exp(i\Delta tHj)\big( \ket{\Phi} + \ket{\psi} \big), \nonumber \\
    f_{4}[j] =&\: \bra{\Phi}\exp(i\Delta tHj)\big( \ket{\Phi} + i\ket{\psi} \big), 
\end{align}
\end{center}
for our states $\ket{\Phi}$ of interest and a secondary state $\ket{\psi}$. We will come back to which choices of $\ket{\psi}$ are appropriate, but we note that preparing their superposition ideally should not require execution of very deep circuits. 

\subsubsection{The noiseless scenario}
\label{sec:noiselessR1}
Let us first address the setting in which one has access to the \textit{noiseless} absolute value measurements as input to the v-PR problem. At the core of the vectorial phase retrieval algorithm is the understanding of when Problem \ref{prob:simpleVPR} has a unique solution. To illustrate when uniqueness is guaranteed, let us introduce the following notions.

\bigbreak
\noindent 
\textit{The DFTs $\mathbf{F}_{1}\in \mathbb{C}^{N}$ and $\mathbf{F}_{2}\in \mathbb{C}^{N}$ 
of the time-series $\mathbf{f}_{1}$ and $\mathbf{f}_{2}$
are spectrally independent if $P_{1}(z) := \sum_{k=0}^{N-1}F_{1}[k]\:z^{k}$ (with $z\in \mathbb{C}$) and $P_{2}(z) := \sum_{k=0}^{N-1}F_{2}[k]\:z^{k}$ (with $z\in\mathbb{C}$) have no common roots in the complex plane.}
\bigbreak

\noindent 
\textit{The DFTs $\mathbf{F}_{1}\in \mathbb{C}^{N}$ and $\mathbf{F}_{2}\in \mathbb{C}^{N}$ 
of the time-series $\mathbf{f}_{1}$ and $\mathbf{f}_{2}$ have well-defined support if they are such that there exist $k_{\min}<k_{\max}\in \{0,1,\ldots,N-1\}$ s.t. $k_{\max}-k_{\min} = \sigma\leq \lfloor N/2 \rfloor$ and $|F_{1}[k]|,|F_{2}[k]|\neq 0$ for $k_{\min}\leq k \leq k_{\max}$ and $|F_{1}[k]| = |F_{2}[k]| = 0$ for $k<k_{\min}$ and $k>k_{\max}$. In other words, they have well-defined support if they are non-zero inside of some interval $(k_{\min},\ldots,k_{\max})$ (of length $\sigma\leq \lfloor N/2 \rfloor$), and zero outside of that interval. }
\bigbreak

\noindent
Especially the latter notion will play a central role in the vectorial phase retrieval algorithm, and in particular in our application of the algorithm to spectral estimation of Hamiltonians. The uniqueness of the solution of the vectorial phase retrieval problem is related to these notions in the following way. 

\bigbreak
\noindent
\textit{The vectorial phase retrieval problem has a unique solution if and only if $\mathbf{F}_{1}$ and $\mathbf{F}_{2}$ are spectrally independent and have well-defined support \cite{VPR}. }
\bigbreak

\noindent
See \cite{VPR} for a proof of this statement. Note that Problem \ref{prob:simpleVPR} has trivial ambiguities. The trivial ambiguities correspond in general to all signals related to $(\mathbf{f}_{1},\mathbf{f}_{2})$ by 
\vspace{-0.7cm}
\begin{center}
\begin{align}
    &\big\{\exp(i\alpha)\:F_{1}[k],\exp(i\alpha)\:F_{2}[k]\big\}_{k=0}^{N-1} \text{ (with $\alpha \in \mathbb{R}$)},\nonumber \\ &\big\{F_{1}[k-k_{0}], F_{2}[k-k_{0}]\big\}_{k=0}^{N-1} \text{ (with $k_{0} \in \mathbb{Z}$)},\nonumber \\ &\big\{F_{1}^{*}[-k],F_{2}^{*}[-k]\big\}_{k=0}^{N-1}, 
\end{align}
\end{center}
or combinations thereof. Solutions to the vectorial phase retrieval problem will only ever be unique up to these trivial ambiguities. 

The vectorial phase retrieval algorithm (presented in \cite{VPR}) is based on the fact that the unique solution to Problem \ref{prob:simpleVPR} can be obtained by solving a convex optimization problem in the ideal noiseless scenario. The optimization procedure optimizes over assignments of phases to the time series measurements $\{|f_{1}[j]|\}$ and $\{|f_{2}[j]|\}$ such that they are consistent with the interference measurements and such that $\mathbf{F}_{1}$ and $\mathbf{F}_{2}$ have the right support size. 

The \textit{input} to the v-PR algorithm consists of $|f_1[j]|^2,\:|f_{2}[j]|^2,\: |f_{3}[j]|^2$ and $|f_{4}[j]|^2$ (at $j = 0,1,\ldots,N-1$), and of an estimate $s$ of the support size $\sigma$ of $\mathbf{F}_{1}$ and $\mathbf{F}_{2}$. Its \textit{output} is given by a vector $\mathbf{y}\in \mathbb{C}^{2N}$, whose entries are the estimates of the phases of $f_1[j]$ and $f_2[j]$ at $j = 0,1,\ldots,N-1$. The algorithm is an optimization procedure that minimizes a cost function over possible assignments of phases $\mathbf{y}\in \mathbb{C}^{2N}$. Let us denote the vector containing the correct phases of $\mathbf{f}_1$ and $\mathbf{f}_2$ at $j = 0,1,\ldots,N-1$ by $\mathbf{x}$. 
The cost function has two components; one component ensures that the assignment of phases is consistent with the interference measurements, and the other component ensures that $\mathbf{F}_{1}$ and $\mathbf{F}_{2}$ are indeed zero outside of the interval of size $s$ (which is an estimate of $\sigma$). These components are denoted by $Q^{(s)}_{\text{support}}(\mathbf{y})$ and $Q_{\text{interference}}(\mathbf{y})$, respectively. Note that due to the trivial ambiguities, minimization of $|F_1[k]|$ and $|F_2[k]|$ outside of their supported interval can be realized by simply minimizing $|F_1[k]|$ and $|F_2[k]|$ at $k=s,s+1,\ldots,N-1$ without loss of generality.

The expression for the support component of the cost function is as follows. 
\vspace{-0.8cm}
\begin{center}
\begin{align}
    Q^{(s)}_{\text{support}}(\mathbf{y}) :=&\: \sum_{k=s}^{N-1} \bigl\lvert F_{1}[k] \bigr\rvert^{2} + \sum_{k=s}^{N-1} \bigl\lvert F_{2}[k] \bigr\rvert^{2} \nonumber \\ =&\: \sum_{k=s}^{N-1} \Bigl\lvert \sum_{j=0}^{N-1}|f_{1}[j]|\: y_{j}\exp\big[-i2\pi jk/N\big]\Bigr\rvert^{2} + \sum_{k=s}^{N-1} \Bigl\lvert \sum_{j=0}^{N-1}|f_{2}[j]|\: y_{N+j}\exp\big[-i2\pi jk/N\big]\Bigr\rvert^{2}.
\end{align}
\end{center}

The expression for the interference component of the cost function is a slightly more involved. Because of the following relations derived from the definitions of $\mathbf{f}_{3}$ and $\mathbf{f}_{4}$ and involving the correct phases $\mathbf{x}$,
\vspace{-0.5cm}
\begin{center}
\begin{align}
    |f_{3}[j]|^{2} =&\: |f_{1}[j] + f_{2}[j]|^{2} = |f_{1}[j]|^{2} + |f_{2}[j]|^{2} + 2|f_{1}[j]|\:|f_{2}[j]|\:\text{Re}\big(x_{j}x^{*}_{N+j}\big),\nonumber \\
    |f_{4}[j]|^{2} =&\: |f_{1}[j] + if_{2}[j]|^{2} = |f_{1}[j]|^{2} + |f_{2}[j]|^{2} + 2|f_{1}[j]|\:|f_{2}[j]|\:\text{Im}\big(x_{j}x^{*}_{N+j}\big),
\end{align}
\end{center}
we have that 
\begin{equation}
    x_{j} = x_{N+j}\: G[j], \quad G[j]:= \frac{|f_{3}[j]|^{2} + i|f_{4}[j]|^{2} - (1+i)\big( |f_{1}[j]|^{2} + |f_{2}[j]|^{2} \big)}{2|f_{1}[j]|\:|f_{2}[j]|}.
\end{equation}
We note that $|G[j]| = 1$ for $j = 0,1,\ldots,N-1$. The second component of the cost function is expressed as follows, and is equal to zero if $\mathbf{y} = \mathbf{x}$.
\begin{equation}
    Q_{\text{interference}}(\mathbf{y}) := \sum_{j=0}^{N-1} \:\bigl\lvert\:y_{j} - y_{N+j}\:G[j]\:\bigr\rvert^{2}. 
\end{equation}

The total cost function is defined simply as the sum of the support component and the interference component. We note that this cost function is non-negative by definition. 
\begin{equation}
    Q^{(s)}(\mathbf{y}) := Q^{(s)}_{\text{support}}(\mathbf{y}) + Q_{\text{interference}}(\mathbf{y}).
\end{equation}
Importantly, it can be expressed as 
\begin{equation}
    Q^{(s)}(\mathbf{y}) := \mathbf{y}^{\dagger}A_{s}^{\dagger}A_{s}\:\mathbf{y},
\end{equation}
where $A_{s}\in \mathbb{C}^{(2(N-s)+N)\times 2N}$ has entries
\small
\begin{equation}
    (A_{s})_{p,q} = 
    \begin{cases}
      |f_{1}[q]|\exp(i2\pi q(s+p)/N), & \text{if $q<N$, $p<N-s$,} \\
      |f_{2}[q]|\exp(i2\pi (q-N)(s+p-(N-s))/N), & \text{if $N\leq q < 2N$, $(N-s)\leq p < 2(N-s)$,} \\ 
      1, & \text{if $q<N$, $p = q + 2(N-s)$,} \\
      -G[q-N], & \text{if $N\leq q<2N$, $p = q-N + 2(N-s)$,} \\ 
      0, & \text{otherwise},
    \end{cases} 
\end{equation}
\normalsize
with $p=0,1,\ldots,(2(N-s)+N) - 1$ and $q = 0,1,\ldots,2N - 1$. 

If there exists an assignment of phases $\mathbf{y}$ such that $Q^{(s)}(\mathbf{y}) = 0$, then we are guaranteed that for that assignment $F_{1}[k] = F_{2}[k] = 0$ at $k=s,s+1,\ldots,N-1$, and $y_{j} = y_{N+j}\:G[j]$ at $j=0,1,\ldots,N-1$. The correct assignment of phases $\mathbf{x}$ is a minimizer of the cost function $Q^{(\sigma)}(\mathbf{y})$. In particular, since the correct assignment of phases attains zero cost on both cost functions (by definition), we have that 
\begin{equation}
    \phi_{\sigma}^{\text{noiseless}} := \hspace{-0.4cm} \min_{\substack{\mathbf{y}\in\mathbb{C}^{2N}\text{ s.t. } \\ |y_{0}| = |y_{1}| = \ldots = |y_{N-1}| = 1}} \hspace{-0.4cm}Q^{(\sigma)}(\mathbf{y}) = Q^{(\sigma)}(\mathbf{x}) = 0,
\label{eq:originaloptproblem2}
\end{equation}
where the optimization is over all vectors $\mathbf{y}\in \mathbb{C}^{N}$ whose entries are phase factors.

We are not, however, a priori guaranteed that $\mathbf{x}$ is the unique minimizer of $Q^{(\sigma)}(\mathbf{y})$, and therefore that solving Eq. \eqref{eq:originaloptproblem2} provides the correct assignment of phases $\mathbf{x}$. However, it was shown in \cite{VPR} that $Q^{(\sigma)}(\mathbf{y}) = 0$ for \textit{any} vector $\mathbf{y}\in \mathbb{C}^{2N}$ (with non-zero 2-norm) iff $\mathbf{y}$ is the exact assignment $\mathbf{x}$. From this fact we conclude that the correct assignment of phases $\mathbf{x}$ can also be obtained by solving the following relaxation of the non-convex optimization problem in Eq. \eqref{eq:originaloptproblem2}.
\begin{equation}
    \chi_{\sigma}^{\text{noiseless}} := \hspace{-0.4cm} \min_{\substack{\mathbf{y}\in\mathbb{C}^{2N}\text{ s.t. } \\ ||\mathbf{y}||_{2} = \sqrt{2N}}} \hspace{-0.2cm}Q^{(\sigma)}(\mathbf{y}) = (R+1)N\:\lambda_{\min}\big( A_{\sigma}^{\dagger}A_{\sigma} \big) = Q^{(\sigma)}(\mathbf{x}) = 0,
\end{equation}
where the optimization is over \textit{all} vectors with a 2-norm equal to $\sqrt{2N}$. In other words, in the noiseless scenario, the correct assignment of phases $\mathbf{x}$ can be obtained by determining the smallest eigenvector of $A_{\sigma}^{\dagger}A_{\sigma}$.

\subsubsection{The noisy scenario}
In practice, one only obtains the noisy versions of the absolute value measurements $|f_1[j]|^2$, $|f_{2}[j]|^2$, $|f_{3}[j]|^2$ and $|f_{4}[j]|^2$, which we respectively denote by $|\tilde{f}_1[j]|^2,\:|\tilde{f}_{2}[j]|^2,\: |\tilde{f}_{3}[j]|^2$ and $|\tilde{f}_{4}[j]|^2$. We denote the correct vector of phases of $\tilde{\mathbf{f}}_1$ and $\tilde{\mathbf{f}}_{2}$ by $\tilde{\mathbf{x}}\in \mathbb{C}^{2N}$. Let us now define the cost function in terms of these noisy absolute value measurements as follows. 
\begin{equation}
    \tilde{Q}^{(s)}(\mathbf{y}) := \tilde{Q}^{(s)}_{\text{support}}(\mathbf{y}) + \tilde{Q}_{\text{interference}}(\mathbf{y}),
\end{equation}
with
\begin{equation}
    \tilde{Q}^{(s)}_{\text{support}}(\mathbf{y}) := \sum_{k=s}^{N-1} \Bigl\lvert \frac{1}{N}\sum_{j=0}^{N-1}|\tilde{f}_{1}[j]|\: y_{j}\exp\big[-i2\pi jk/n\big]\Bigr\rvert^{2} + \sum_{k=s}^{N-1} \Bigl\lvert \frac{1}{N}\sum_{j=0}^{N-1}|\tilde{f}_{2}[j]|\: y_{N+j}\exp\big[-i2\pi jk/n\big]\Bigr\rvert^{2},
\end{equation}
and
\begin{equation}
    \tilde{Q}_{\text{interference}}(\mathbf{y}) := \sum_{j=0}^{N-1} \:\bigl\lvert\:y_{j} - y_{N+j}\:\tilde{G}[j]/|\tilde{G}[j]|\:\bigr\rvert^{2}, \:\:\:\: \tilde{G}[j]:= \frac{|\tilde{f}_{3}[j]|^{2} + i|\tilde{f}_{4}[j]|^{2} - (1+i)\big( |\tilde{f}_{1}[j]|^{2} + |\tilde{f}_{2}[j]|^{2} \big)}{2|\tilde{f}_{1}[j]|\:|\tilde{f}_{2}[j]|},
\end{equation}
where we stress that $\tilde{G}[j]$ does not necessarily have unit absolute value. 

Again, the cost function can be expressed as 
\begin{equation}
    Q^{(s)}(\mathbf{y}) := \mathbf{y}^{\dagger}\tilde{A}_{s}^{\dagger}\tilde{A}_{s}\:\mathbf{y},
\end{equation}
where $\tilde{A}_{s}\in \mathbb{C}^{(2(N-s)+N)\times 2N}$ has entries
\small
\begin{equation}
    (\tilde{A}_{s})_{p,q} = 
    \begin{cases}
      |\tilde{f}_{1}[q]|\exp(i2\pi q(s+p)/N), & \text{if $q<N$, $p<N-s$,} \\
      |\tilde{f}_{2}[q]|\exp(i2\pi (q-N)(s+p-(N-s))/N), & \text{if $N\leq q < 2N$, $(N-s)\leq p < 2(N-s)$,} \\ 
      1, & \text{if $q<N$, $p = q + 2(N-s)$,} \\
      -\tilde{G}[q-N]/|\tilde{G}[q-N]|, & \text{if $N\leq q<2N$, $p = q-N + 2(N-s)$,} \\ 
      0, & \text{otherwise},
    \end{cases} 
\end{equation}
\normalsize
with $p=0,1,\ldots,(2(N-s)+N) - 1$ and $q = 0,1,\ldots,2N - 1$. 

In the noisy setting, there is no guarantee that there is an assignment of phases $\mathbf{y}$ for which $\tilde{Q}^{(s)}(\mathbf{y}) = 0$, but the optimal assignment of phases could be found by solving the non-convex optimization problem 
\begin{equation}
    \phi_{s}^{\text{noisy}} := \hspace{-0.4cm} \min_{\substack{\mathbf{y}\in\mathbb{C}^{2N}\text{ s.t. } \\ |y_{0}| = |y_{1}| = \ldots = |y_{N-1}| = 1}} \hspace{-0.4cm}\tilde{Q}^{(s)}(\mathbf{y}),
\label{eq:noisyoriginaloptproblem2}
\end{equation}
at $s = \sigma$. In general, we now have that $\phi_{s}^{\text{noisy}} > 0$. The correct assignment $\tilde{\mathbf{x}}$ is not guaranteed to be the unique minimizer of $\tilde{Q}^{(\sigma)}(\mathbf{y})$. However, for relatively small noise magnitudes, any assignment of phases for which $\tilde{Q}^{(\sigma)}(\mathbf{y})$ is minimized tends to be close to the correct assignment $\tilde{\mathbf{x}}$. Like in the noiseless scenario, we will consider the following relaxation of the problem in Eq. \eqref{eq:noisyoriginaloptproblem2} to an eigenvalue problem.
\begin{equation}
    \chi_{\sigma}^{\text{noisy}} := \hspace{-0.4cm} \min_{\substack{\mathbf{y}\in\mathbb{C}^{2N}\text{ s.t. } \\ ||\mathbf{y}||_{2} = \sqrt{2N}}} \hspace{-0.2cm}\tilde{Q}^{(\sigma)}(\mathbf{y}) = 2N\:\lambda_{\min}\big( \tilde{A}_{\sigma}^{\dagger}\tilde{A}_{\sigma} \big),
\label{eq:relaxationnoisy2}
\end{equation}
where we note that again, generally, $\chi_{\sigma}^{\text{noisy}} > 0$. For relatively small noise magnitudes, the optimal assignment obtained by solving Eq. \eqref{eq:relaxationnoisy2} (which we denote by $\mathbf{y}_{\min}$) will be close to the correct assignment $\tilde{\mathbf{x}}$.

\bigbreak 
Before discussing the details of the algorithm, let us consider a generalization that we introduce in this work. This generalization consists of the inclusion of \textit{multiple} time series that are used for interference, as opposed to just a single one. As we shall demonstrate, this leads to improved performance of the algorithm in the presence of noise.

\subsection{Vectorial phase retrieval using multiple interference signals}
\label{sec:multintvpr}

In this work, we consider the scenario in which one employs \textit{multiple} signals $\mathbf{f}_{2}^{(r)} \in \mathbb{C}^{N}$ (with $r=1,2,\ldots,R$) with which the target time series $\mathbf{f}_{1}$ interferes. As we will demonstrate in Section \ref{sec:vecpr_numerics}, we find that the performance of vectorial phase retrieval improves when multiple interference signals are employed in the noisy scenario. This multi-interference vectorial phase retrieval problem is formulated as follows. 
\begin{problem}
    \normalfont{[Multi-interference vectorial phase retrieval]} Given measurements of the absolute values 
    \begin{equation}
        |f_1[j]|^2,\:\: \Big\{|f_{2}^{(r)}[j]|^2,\: |f_{3}^{(r)}[j]|^2:=|f_1[j]+f_{2}^{(r)}[j]|^2,\: |f_{4}^{(r)}[j]|^2:=|f_1[j]+if_{2}^{(r)}[j]|^2\Big\}_{r=1}^{R},
    \end{equation}
    at times $j=0,1,\ldots,N-1$, determine 
    \begin{equation}
        F_1[k] := \sum_{j=0}^{N-1}f_1[j]\exp\big[-i2\pi jk/N\big], \:\: \Big\{ F_{2}^{(r)}[k] := \sum_{j=0}^{N-1}f_{2}^{(r)}[j]\exp\big[-i2\pi jk/N\big] \Big\}_{r=1}^{R},
    \end{equation}
    at $k=0,1,\ldots,N-1$.
\label{prob:VPR}
\end{problem}
\noindent 
The trivial ambiguities of Problem \ref{prob:VPR} are equivalent to those of Problem \ref{prob:simpleVPR}, involving \textit{all} $(R+1)$ signals $\mathbf{F}_{1},\mathbf{F}_{2}^{(1)},\ldots,\mathbf{F}_{2}^{(R)}$ (instead of just $\mathbf{F}_{1},\mathbf{F}_{2}$).

\bigbreak
We will discuss how the multi-interference v-PR algorithm can be implemented to (approximately) solve Problem \ref{prob:VPR}. But let us first discuss how the control-free quantum phase estimation routine fits into the multi-interference v-PR framework. We take 
\begin{equation}
f_{1}[j] = \bra{\Phi} \exp(iH\Delta t\:j)\ket{\Phi} \text{ and } \Big\{ f_{2}^{(r)}[j] = \bra{\Phi} \exp(iH\Delta t\:j)\ket{\psi_{r}} \Big\}_{r=1}^{R},
\end{equation}
such that 
\begin{multline}
\Big\{ f_{3}^{(r)}[j] = f_{1}[j] + f_{2}^{(r)}[j] = \bra{\Phi} \exp(iH\Delta t\:j)\big(\ket{\Phi} + \ket{\psi_{r}}\big) \text{ and } \\ f_{4}^{(r)}[j] = f_{1}[j] + if_{2}^{(r)}[j] = \bra{\Phi} \exp(iH\Delta t\:j)\big(\ket{\Phi} + i\ket{\psi_{r}}\big) \Big\}_{r=1}^{R}.
\end{multline}
Hence by preparing states $\ket{\Phi}$ and $\big\{\ket{\psi_{r}}$, $1/\sqrt{2}\big(\ket{\Phi} + \ket{\psi_{r}}\big)$, $1/\sqrt{2}\big(\ket{\Phi} + i\ket{\psi_{r}}\big)\big\}_{r=1}^{R}$, then time evolving \textit{each} of the states under Hamiltonian $H$ for a time $\Delta t\:j$, and measuring the overlap with $\ket{\Phi}$, we can estimate $|f_{1}[j]|$ and $\big\{|f_{2}^{(r)}[j]|$, $|f_{3}^{(r)}[j]|$, $|f_{4}^{(r)}[j]|\big\}_{r=1}^{R}$ at $j=0,1,\ldots,N-1$. Of course, we do not want the states $\ket{\Phi}$ and $\{\ket{\psi_{r}}\}_{r=1}^{R}$ to be such that preparing superpositions of $\ket{\Phi}$ with any $\ket{\psi_{r}}$ requires applying circuits of very large depth.

As discussed before, one only ever obtains noisy versions of the absolute value measurements $|f_1[j]|^2$, $\big\{|f_{2}^{(r)}[j]|^2$, $|f_{3}^{(r)}[j]|^2$, $|f_{4}^{(r)}[j]|^2\big\}_{r=1}^{R}$ in practice, which we denote by $|\tilde{f}_1[j]|^2$, $\big\{|\tilde{f}_{2}^{(r)}[j]|^2$, $|\tilde{f}_{3}^{(r)}[j]|^2$, $|\tilde{f}_{4}^{(r)}[j]|^2\big\}_{r=1}^{R}$. The multi-interference v-PR algorithm takes as \textit{input} these noisy absolute value measurements, and an estimate $s$ of $\sigma$ (with $\sigma$ denoting the support size of $\mathbf{F}_{1},\mathbf{F}_{2}^{(1)},\ldots,\mathbf{F}_{2}^{(R)}$). The \textit{output} of the algorithm is a vector $\mathbf{y} \in \mathbb{C}^{(R+1)N}$ whose entries are estimates of the phases of $f_{1}[j]$, $f_{2}^{(1)}[j]$, $f_{2}^{(2)}[j]$, $\ldots$, $f_{2}^{(R)}[j]$ at $j=0,1,\ldots,N-1$. The algorithm again consists of the minimization of a cost function over all these assignments of phases to the absolute value measurements $\{|\tilde{f}_1[j]|\}, \: \{|\tilde{f}_2^{(1)}[j]|\},\: \ldots,\:\{|\tilde{f}_2^{(R)}[j]|\}$. We denote by $\tilde{\mathbf{x}}\in \mathbb{C}^{(R+1)N}$ the correct phases of $f_{1}[j]$, $f_{2}^{(1)}[j]$, $f_{2}^{(2)}[j]$, $\ldots$, $f_{2}^{(R)}[j]$ at $j=0,1,\ldots,N-1$.

The cost function again consists of a support component $\tilde{Q}^{(s)}_{\text{support}}(\mathbf{y})$ and an interference component $\tilde{Q}_{\text{interference}}(\mathbf{y})$. The support component of the cost function can be written as 
\begin{equation}
    \tilde{Q}^{(s)}_{\text{support}}(\mathbf{y}) := \sum_{k=s}^{N-1} \Bigl\lvert \sum_{j=0}^{N-1}|\tilde{f}_{1}[j]|\: y_{j}\exp\big[-i2\pi jk/n\big]\Bigr\rvert^{2} + \sum_{r=1}^{R}\Bigg[ \sum_{k=s}^{N-1} \Bigl\lvert \sum_{j=0}^{N-1}|\tilde{f}_{2}^{(r)}[j]|\: y^{(r)}_{j}\exp\big[-i2\pi jk/n\big]\Bigr\rvert^{2} \Bigg],
\end{equation}
which is equivalent to the support component defined in the single-interference-signal framework, except that now it constrains the support of $R+1$ signals. The interference component $\tilde{Q}_{\text{interference}}(\mathbf{y})$, which is again an extension of the single-interference-signal interference component to $R+1$ signals, can be defined as follows.

\begin{equation}
    \tilde{Q}_{\text{interference}}(\mathbf{y}) := \sum_{r=1}^{R}\bigg[\sum_{j=0}^{N-1} \:\Bigl\lvert\:y_{j} - y^{(r)}_{j}\tilde{G}_{r}[j]/|\tilde{G}_{r}[j]|\:\Bigr\rvert^{2} \bigg], 
\end{equation}
where the quantities $\tilde{G}_{r}[j]$ are defined at $j=0,1,\ldots,N-1$ for $r = 1,2,\ldots,R$ as
\begin{equation}
    \tilde{G}_{r}[j]:= \frac{|\tilde{f}_{3}^{(r)}[j]|^{2} + i|\tilde{f}_{4}^{(r)}[j]|^{2} - (1+i)\big( |\tilde{f}_{1}[j]|^{2} + |\tilde{f}_{2}^{(r)}[j]|^{2} \big)}{2|\tilde{f}_{1}[j]|\:|\tilde{f}_{2}^{(r)}[j]|}.
\end{equation}

The total cost function
\begin{equation}
    \tilde{Q}^{(s)}(\mathbf{y}) := \tilde{Q}^{(s)}_{\text{support}}(\mathbf{y}) + \tilde{Q}_{\text{interference}}(\mathbf{y})
\end{equation}
can be expressed as 
\begin{equation}
    \tilde{Q}^{(s)}(\mathbf{y}) := \mathbf{y}^{\dagger}\tilde{A}_{s}^{\dagger}\tilde{A}_{s}\:\mathbf{y},
\end{equation}
where $\tilde{A}_{s,R}\in \mathbb{C}^{((R+1)(N-s)+RN)\times (R+1)N}$ has entries
\footnotesize
\begin{equation}
    (A_{s})_{p,q} = 
    \begin{cases}
      |\tilde{f}_{1}[q]|\exp(i2\pi q(s+p)/N), & \text{if $q<N$, $p<N-s$,} \\
      |\tilde{f}_{2}[q]|\exp(i2\pi (q-N)(s+p-(N-s))/N), & \text{if $N\leq q < 2N$, $(N-s)\leq p < 2(N-s)$,} \\ 
      \:\vdots & \\
      |\tilde{f}_{R+1}[q]|\exp(i2\pi (q-RN)(s+p-R(N-s))/N), & \text{if $RN\leq q < (R+1)N$, $R(N-s)\leq p < (R+1)(N-s)$,} \\ 
      1, & \text{if $q<N$, $p = q + (R+1)(N-s)$,} \\
      -\tilde{G}_{2}[q-N]/|\tilde{G}_{2}[q-N]|, & \text{if $N\leq q<2N$, $p = q-N + (R+1)(N-s)$,} \\
      1, & \text{if $q<N$, $p = q + (R+1)(N-s)+N$,} \\
      -\tilde{G}_{3}[q-2N]/|\tilde{G}_{3}[q-2N]|, & \text{if $2N\leq q<3N$, $p = q-N + (R+1)(N-s)$,} \\ 
      \:\vdots & \\
      1, & \text{if $q<N$, $p = q +(R+1)(N-s)+(R-1)N$,} \\
      -\tilde{G}_{R+1}[q-RN]/|\tilde{G}_{R+1}[q-RN]|, & \text{if $RN\leq q<(R+1)N$, $p = q-N + (R+1)(N-s)$,} \\ 
      0, & \text{otherwise},
    \end{cases} 
\end{equation}
\normalsize
with $p=0,1,\ldots,((R+1)(N-s)+RN) - 1$ and $q = 0,1,\ldots,(R+1)N - 1$.

The optimal assignment of phases could be found by solving the non-convex optimization problem 
\begin{equation}
    \Phi_{\sigma}^{\text{noisy}} := \hspace{-0.4cm} \min_{\substack{\mathbf{y}\in\mathbb{C}^{(R+1)N}\text{ s.t. } \\ |y_{0}| = |y_{1}| = \ldots = |y^{(R)}_{N-1}| = 1}} \hspace{-0.4cm}\tilde{Q}^{(\sigma)}(\mathbf{y}).
\label{eq:noisyoriginaloptproblem}
\end{equation}
In general, we have that $\Phi_{\sigma}^{\text{noisy}} > 0$. The correct assignment $\mathbf{\tilde{x}}$ is again not guaranteed to be a (unique) minimizer of $\tilde{Q}^{(\sigma)}(\mathbf{y})$. However, for relatively small noise magnitudes, any assignment of phases for which $\tilde{Q}^{(\sigma)}(\mathbf{y})$ is minimized tends to be close to the correct assignment $\mathbf{\tilde{x}}$. 

We again relax the problem in Eq. \eqref{eq:noisyoriginaloptproblem} to an eigenvalue problem as follows. 
\begin{equation}
    X_{\sigma}^{\text{noisy}} := \hspace{-0.4cm} \min_{\substack{\mathbf{y}\in\mathbb{C}^{(R+1)N}\text{ s.t. } \\ ||\mathbf{y}||_{2} = \sqrt{(R+1)N}}} \hspace{-0.2cm}Q^{(\sigma)}(\mathbf{y}) = (R+1)N\:\lambda_{\min}\big( \tilde{A}_{\sigma}^{\dagger}\tilde{A}_{\sigma} \big),
\label{eq:relaxationnoisy}
\end{equation}
where we note that again, generally, $X_{\sigma}^{\text{noisy}} > 0$. As we shall demonstrate in Section \ref{sec:vecpr_numerics}, for relatively small noise magnitudes, the optimal assignment $\mathbf{y}_{\min}$ obtained by solving Eq. \eqref{eq:relaxationnoisy} will be close to the correct assignment $\tilde{\mathbf{x}}$. Moreover, we will show that for a given noise magnitude, the assignment of phases obtained by solving Eq. \eqref{eq:relaxationnoisy} will typically become a better approximation for $\tilde{\mathbf{x}}$ as $R$ increases. 

\begin{comment}
Since the optimization problems in Eq. \eqref{eq:relaxations} are relaxations of those in Eqs. \eqref{eq:originaloptproblem} and \eqref{eq:noisyoriginaloptproblem}, we have that
\begin{equation}
    \chi_{s,R}^{\text{noiseless}} \leq \phi_{s,R}^{\text{noiseless}} \quad \text{and} \quad \chi_{s,R}^{\text{noisy}} \leq \phi_{s,R}^{\text{noisy}}.
\end{equation}
Because of Eq. \ref{eq:originaloptproblem}, we conclude that $\chi_{s,R}^{\text{noiseless}} = 0$ and since $Q^{(s,R)}(\mathbf{y}) = 0$ (for \textit{any} $\mathbf{y}\in \mathbb{C}^{(R+1)n}$) iff $\mathbf{y} = \mathbf{x}$, the smallest eigenvector (when normalized s.t. its 2-norm is $\sqrt{(R+1)n}$) of $A_{s,R}^{\dagger}A_{s,R}$ is equal to $\mathbf{x}$. 
\end{comment}

\bigbreak
We note that the optimal assignment $\mathbf{y}_{\min}$ obtained by solving Eq. \eqref{eq:relaxationnoisy} is not necessarily an assignment of phases. Although we ensure that $||\mathbf{y}_{\min}||_{2} = \sqrt{(R+1)N}$, each entry can have absolute value different from one. One could consider rounding $\mathbf{y}_{\min}$ to an assignment of phases by dividing each entry by its absolute value. We do not perform this rounding as we find it does not \textit{necessarily} improve the approximation to $\tilde{\mathbf{x}}$ in the scenarios that we consider. 

In Section \ref{sec:vecpr_numerics}, we will demonstrate numerically how the v-PR algorithm can be used to perform spectral estimation for instances of the FH Hamiltonian, and discuss its performance in various settings. 

%\begin{equation}
%    f_{1,\text{approx}} := \text{d-}\mathcal{FT}\Big[ \big( \: |\tilde{F}_{1}(0)|\:y_{\min,0}^{(1)},\: |\tilde{F}_{1}(1)|\:y_{\min,1}^{(1)},\: \ldots, |\tilde{F}_{1}(n-1)|\:y_{\min, \:n-1}^{(1)} \: \big) \Big]. 
%\end{equation}

\begin{comment}

%We will not consider here the error to exact time dynamics this simulation introduces as the results offered in this paper are not prescriptive, but rather allow for demonstration of the principles of phase retrieval being applicable to dynamics expected from Fermi-Hubbard model.
\end{comment}

\subsection{Vectorial phase retrieval algorithm for spectral estimation of Hamiltonians}
\label{sec:vecpr_numerics}

In this section, we numerically investigate the performance of the vectorial phase retrieval algorithm in the context of spectral estimation for Hamiltonians. The signals encountered in this context do not perfectly fit within the framework of v-PR. In particular, due to the fact that the (absolute values of) the time series are only measured for a finite measurement time, the signals $\mathbf{F}_{1},\: \mathbf{F}_{2}^{(1)},\: \ldots,\: \mathbf{F}_{2}^{(R)}$ do not necessarily have well-defined support. Typically, $|F_{1}[k]|,\: |F_{2}^{(1)}[k]|,\: \ldots,\: |F_{2}^{(R)}[k]|$ decay away from the supported eigenvalues, rather than being equal to zero exactly outside of some frequency interval. The more the signals suffer from this effect, the more ill-defined their support is becomes. As a result, a good performance of the vectorial phase retrieval algorithm cannot be guaranteed without further considerations. 

\begin{figure}[H]
        \centering
        \includegraphics[width=.65\linewidth]{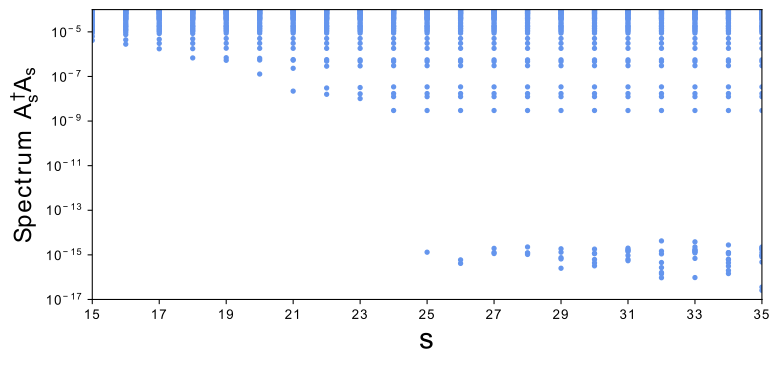}
        \caption{Spectrum of $A_{s}^{\dagger}A_{s}$ for $R=1$ in the noiseless setting, as a function of $s$. The time series amplitudes used to construct $A_{s}^{\dagger}A_{s}$ correspond to evolution under a $1\times 5$ (i.e., $10$-mode) Fermi-Hubbard Hamiltonian for which $|\tau/U| = 1/4$. We have taken $N = 300$. The signals $f_{1}$ and $f_{2}$ were artificially processed to ensure that they have well-defined support (here, $\sigma = 25$).}
        \label{fig:platfigure}
\end{figure}

To set the stage, we first provide an example for which the v-PR input data is obtained from time evolution of some state under the FH Hamiltonian, where we have artificially ensured that the input data fits exactly in the v-PR framework. Namely, we have altered the input data in such a way that $\mathbf{F}_{1},\: \mathbf{F}_{2}^{(1)}$ (we take $R=1$ here) are exactly equal to zero outside of an interval of size $\sigma = 25$. Note that to do this, one generally needs access to the full time series and not just their absolute values (so this cannot be done in practice). The data we have used here is noiseless data. Based on the discussion in Section \ref{sec:noiselessR1}, we expect the matrix $A_{s}^{\dagger}A_{s}$ to have a \textit{single} zero eigenvalue at $s = \sigma$, and its smallest eigenvector to correspond to the exact assignment of phases. Figure \ref{fig:platfigure} depicts the spectrum of $A_{s}^{\dagger}A_{s}$ as a function of $s$. Indeed, its smallest eigenvalue is non-zero for $s < \sigma$ and (numerically) zero at $s = \sigma$. For $s>\sigma$, there are several zero-eigenvalue eigenvectors, with the correct solution lying in the span of those eigenvectors. In the noisy setting, there will generally be \textit{no} zero-eigenvalue solutions. However, for sufficiently small noise magnitudes, there will still be a drop in the value of the smallest eigenvalue of $A_{s}^{\dagger}A_{s}$ around $\sigma$. 
Our investigations suggest that:
\begin{enumerate}
    \item The two smallest eigenvalues are a proxy for the quality of the retrieval of phases. That is, the smaller the smallest eigenvalue, the smaller the value of the two residuals eq.(\ref{eq:Cost-int}) and (\ref{eq:Cost-support}) above. 
    \item A large gap opening between the two smallest eigenvalues suggests a larger overlap between the smallest eigenvector and the true assignment of phases. This being motivated by the fact that in the ideal noiseless case the solution is unique. 
\end{enumerate}

In the remainder of this section, we will discuss the performance of vectorial phase retrieval for input data obtained from time evolution under FH Hamiltonians. The procedure that is implemented is given as a pseudocode below. As discussed, this input data does not perfectly fit into the v-PR framework. Nevertheless, the v-PR algorithm performs well in performing phase retrieval in the scenarios that we consider. 

\begin{figure}[H]
\centering
\begin{algorithmic}[1]
    \Procedure {v-PR}{$\:|\tilde{f}_{1}[j]|$, $|\tilde{f}_{2}^{(r)}[j]|$, $|\tilde{f}_{3}^{(r)}[j]|$, $|\tilde{f}_{4}^{(r)}[j]|$ for $r=1,2,\ldots,R$, at $j=0,1,\ldots,N-1$.}
    \For {s = 0, 1, \ldots, N-1}
    \State Evaluate $G_{r}[j]$ for $r=1,2,\ldots,R$, at $j=0,1,\ldots,N-1$.
    \State Obtain $\tilde{A}_{s}^{\dagger}\tilde{A}_{s} \in \mathbb{C}^{(R+1)N\times (R+1)N}$.
    \State Evaluate the two smallest eigenvalues of $\tilde{A}_{s}^{\dagger}\tilde{A}_{s}$ (with the smallest denoted by $\lambda_{\min}
    (\tilde{A}_{s}^{\dagger}\tilde{A}_{s})$).
    \EndFor
    \State Obtain optimal $s^{*}$ (heuristically).
    \State Evaluate $\mathbf{y}_{\min}\in \mathbb{C}^{(R+1)N}$ s.t. $\mathbf{y}_{\min}^{\dagger}\tilde{A}_{s^{*}}^{\dagger}\tilde{A}_{s^{*}}\mathbf{y}_{\min} = (R+1)N\:\lambda_{\min}(\tilde{A}_{s^{*}}^{\dagger}\tilde{A}_{s^{*}})$.
    \State Output $\{\:|f_{1}[j]|\:(y_{\min})_{j}\:\}_{j=0}^{N-1}$.
    \EndProcedure
\end{algorithmic}
\end{figure}

\begin{comment}
In the remainder of this section
\end{comment}

\bigbreak
\noindent
\textbf{Interplay between $R$ and the noise resilience of the vectorial phase retrieval algorithm:}
.

\begin{figure}[H]
        \subfloat[Noiseless.]{%
            \includegraphics[width=.593\linewidth]{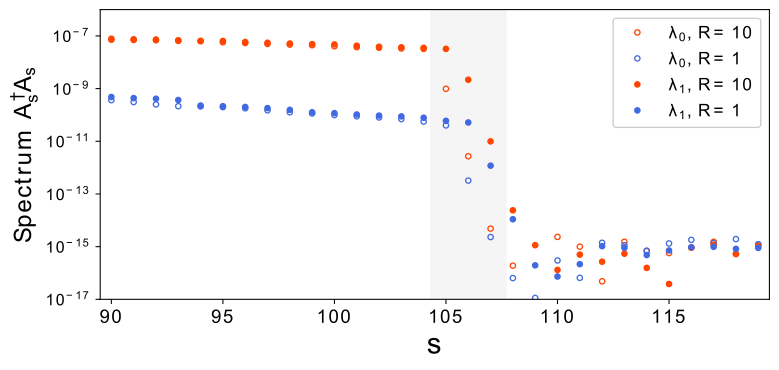}%
            \label{fig:multRa}%
        }\hfill
        \subfloat[Noiseless, $s=105$.]{%
            \vspace{-0.05cm}
            \includegraphics[width=.40\linewidth]{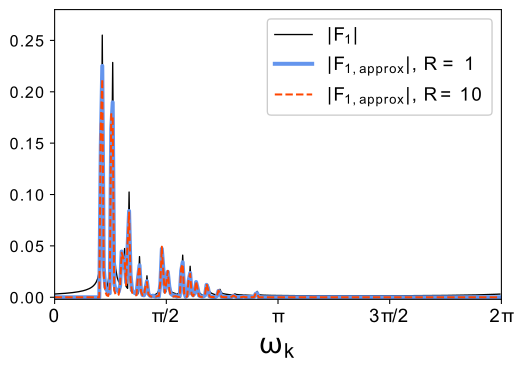}%
            \label{fig:multRb}%
        }\\
        \subfloat[$N_{\text{samples}} = 10^{6}$.]{%
            \includegraphics[width=.593\linewidth]{spectrum_R_noise_resilience_newnoisy.png}%
            \label{fig:multRc}%
        }\hfill
        \subfloat[$N_{\text{samples}} = 10^{6}$, $s=105$.]{%
            \vspace{-0.05cm}
            \includegraphics[width=.40\linewidth]{f_approx_R_noise_resilience.png}%
            \label{fig:multRd}%
        }
        \caption{The smallest two eigenvalues of $A_{s}^{\dagger}A_{s}$ for $R=1$ and $R=10$ in the noiseless setting (see (a)), and for $N_{\text{samples}} = 10^{6}$ (see (c)). The eigenvalues are given as a function of $s$. Approximations to $|F_{1}|$ at fixed $s=105$ for $R=1$ and $R=10$ are depicted in the noiseless setting in (b) and for $N_{\text{samples}} = 10^{6}$ in (d). The time series amplitudes used to construct $A_{s}^{\dagger}A_{s}$ correspond to evolution under a $1\times 5$ (i.e., $10$-mode) Fermi-Hubbard Hamiltonian for which $|\tau/U| = 1/4$. }
        \label{fig:multR}
\end{figure}

\noindent
Figure \ref{fig:multR} illustrates the difference in noise resilience of the v-PR algorithm between the $R=1$ and the $R=10$ scenarios. This difference materializes as follows: In the noisy scenario, there is a sharp drop in the smallest eigenvalue of $\tilde{A}_{s}^{\dagger}\tilde{A}_{s}$ as a function of $s$ for $R = 10$, which is absent for $R=1$. This can be seen in Figure \ref{fig:multR}c. Our numerical investigations suggest that this sharp drop is a signature of an accurate retrieval of phases at that value of $s$. As can be seen in Figure \ref{fig:multR}d, indeed the estimate of $\mathbf{f}_{1}$ for $R=10$ is more accurate than for $R=1$. Figure \ref{fig:multR}b suggests that choosing larger $R$ does not influence the accuracy significantly in the noiseless scenario.

\bigbreak
\noindent
\textbf{Performance of the vectorial phase retrieval algorithm for shallower time evolutions:}

\noindent
The fact that $\mathbf{F}_{1},\: \mathbf{F}_{2}^{(1)},\: \ldots,\: \mathbf{F}_{2}^{(R)}$ do not have well-defined support is the most prominent reason that the v-PR input data in the current setting does not fit perfectly into the framework of v-PR. The support of these signals typically becomes more ill-defined (i.e., $\mathbf{F}_{1},\: \mathbf{F}_{2}^{(1)},\: \ldots,\: \mathbf{F}_{2}^{(R)}$ decay more slowly away from the supported frequencies) as the total time evolution becomes smaller (due to spectral leakage). Since the circuit depths are limited in our near-term scenario, it is of interest to see how the v-PR algorithm fares for shallower time evolutions. 

\begin{figure}[H]
        \subfloat[$R=10$.]{%
            \includegraphics[width=.59\linewidth]{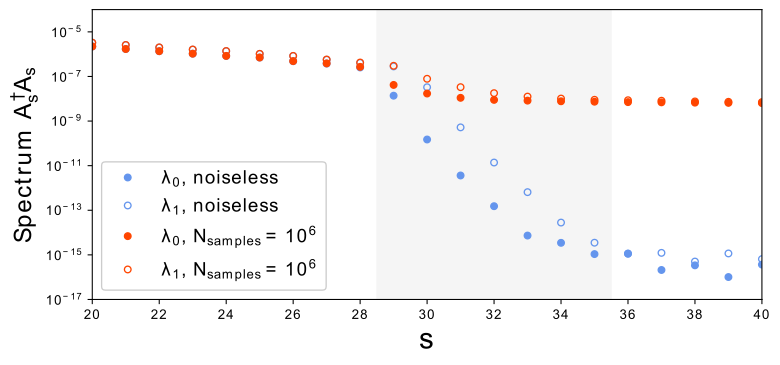}%
            \label{fig:badsupporta}%
        }\hfill
        \subfloat[$R=10$, $s = 29$.]{%
            \vspace{-0.00cm}
            \includegraphics[width=.41\linewidth]{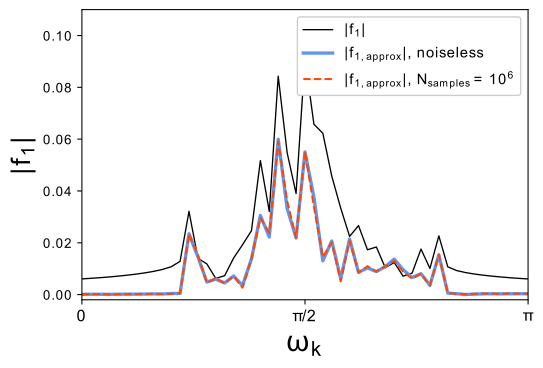}%
            \label{fig:badsupportb}%
        }
        \caption{The smallest two eigenvalues of $A_{s}^{\dagger}A_{s}$ for $R=10$ in the noiseless and noisy settings (see (a)). The eigenvalues are given as a function of $s$. Approximations to $|F_{1}|$ at fixed $s=29$ for $R=10$ are depicted in the noiseless setting and for $N_{\text{samples}} = 10^{6}$ in (b). The time series amplitudes used to construct $A_{s}^{\dagger}A_{s}$ correspond to evolution under a $1\times 5$ (i.e., $10$-mode) Fermi-Hubbard Hamiltonian for which $|\tau/U| = 1/4$, with a relatively small total time evolution $T$. This small total time evolution leads to ill-defined support of $|F_{1}|$. Note that the input state $\ket{\Phi}$ is chosen differently than in Figure \ref{fig:multR}. }
        \label{fig:badsupport}
\end{figure}

Figure \ref{fig:badsupporta} depicts the smallest two eigenvalues $A_{s}^{\dagger}A_{s}$ as a function of $s$. Clearly, in the noiseless setting, $\lambda_{\min}\big(A_{s}^{\dagger}A_{s}\big)$ decays to zero comparatively slowly, which is a result the signals having ill-defined support. We find that the quality of the approximation $|F_{1,\text{approx}}|$ to $|F_{1}|$ is highest at the value of $s$ at which $\lambda_{\min}\big(A_{s}^{\dagger}A_{s}\big)$ first starts to decay, i.e., $s = 29$ for Figure \ref{fig:badsupport}. Figure \ref{fig:badsupportb} depicts $|F_{1,\text{approx}}|$ and in the noiseless scenario, and for $N_{\text{samples}} = 10^{6}$. Remarkably, although the approximation of $|F_{1,\text{approx}}|$ to $|F_{1}|$ is worse (even in the noiseless scenario) than in Figure \ref{fig:multR}, the approximation retains its resilience against the addition of sampling noise. 

\bigbreak
\noindent
\textbf{To round, or not to round:}

\noindent
At the end of Section \ref{sec:multintvpr}, we briefly commented on whether it is beneficial to round the smallest eigenvector $\mathbf{y}_{\min}$ (with 2-norm $\sqrt{(R+1)N}$) of the matrix $\tilde{A}_{s}^{\dagger}\tilde{A}_{s}$ to a vector whose entries are phase factors, by dividing each entry of $\mathbf{y}_{\min}$ by its absolute value. Our numerical investigations suggest that using the entries of this rounded vector to reconstruct the time series does not \textit{necessarily} lead to a more accurate reconstruction. In terms of the reconstructed spectra $F$ (see Figure \ref{fig:toroundornottoround} below, corresponding to the scenario in Figure \ref{fig:comparaison}a), we note the following differences. The unrounded spectrum recovers the peak locations relatively well. The rounded spectrum performs better at recovering the absolute values $|F|$. However, the latter also seems to recover some peaks for frequencies at which $F$ is not actually supported. Whether rounding is beneficial thus depends on which features one wishes to recover. To quantify the error in the particular case of Figure \ref{fig:toroundornottoround}, we note that the 1-norm (normalized by $N$) of the vector with entries $|F_{\text{reconstructed}}[k]| - |F_{\text{exact}}[k]|$ is $0.0016$ without rounding and $0.0013$ with rounding.

\begin{figure}[H]
        \subfloat[Without rounding.]{%
            \includegraphics[width=.49\linewidth]{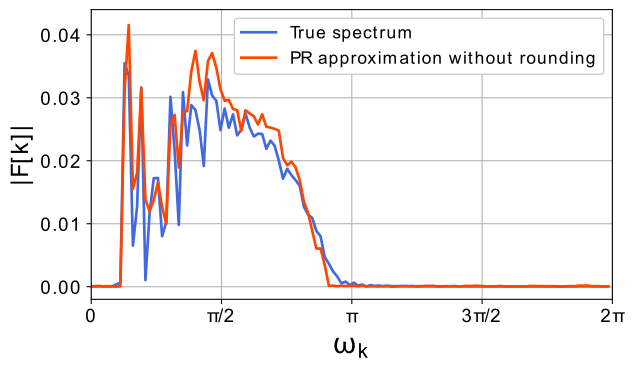}%
            \label{fig:badsupporta}%
        }\hfill
        \subfloat[With rounding.]{%
            \vspace{-0.00cm}
            \includegraphics[width=.49\linewidth]{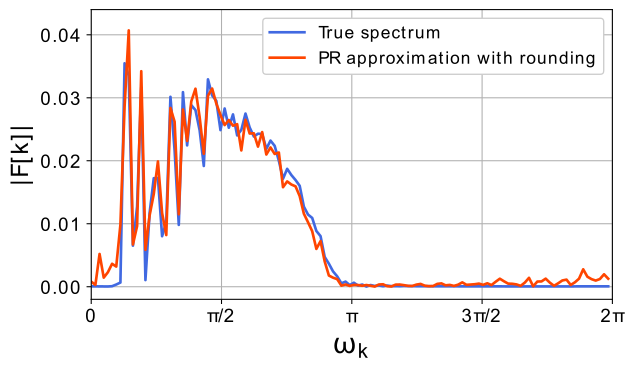}%
            \label{fig:badsupportb}%
        }
        \caption{Comparison between the unrounded (a) and rounded (b) solutions obtained through vectorial phase retrieval. The scenario used to obtain the figures is the same as that of Figure \ref{fig:comparaison}. }
        \label{fig:toroundornottoround}
\end{figure}

\section{Appendix C: 2D Phase Retrieval}
\label{sec:2DPR}
In what follows, we will review the $2$D phase retrieval problem, the hybrid input-output algorithm for solving this problem and then our main contribution. We introduce a procedure for embedding the phase retrieval problem relevant to quantum phase estimation into a $2$D time-series which we show can be guaranteed to be both real and positive. 
Finally, we demonstrate  numerically on a Fermi-Hubbard model example that the HIO algorithm succeeds to retrieve time-series of enough quality to recover a good quality spectrum, showing that the algorithm is resilient to sample noise. In order to keep the analysis simple we just implement a recovery relying on a DFT of the time-series. However, we note that the discrete Fourier transform is not necessarily the most accurate reconstruction of the spectrum of a Hamiltonian. Typically, classical post-processing techniques like the one developed in
\cite{Somma_2019} 
can be used to obtain a more useful description of the spectrum.
\subsection{The 2D Phase Retrieval Problem}
In two-dimensional phase retrieval, one now attempts to solve the following (potentially ambiguous) problem. 
\begin{problem}
	\normalfont{[2D phase retrieval]} Given measurements of the \emph{absolute values}\footnote{The absolute value here means the absolute value of every element $f[j,l]$ of the $N \times M$ matrix $f$.} of a signal $f[j,l]$ determine the discrete Fourier transform $F[k,m]$ of that signal (see Eq. \ref{eq:2d-dft}), up to the trivial ambiguities of global phases, shifts and conjugate reflections.
    We have the following conventions
\begin{align}
	F[k, m] &:= \sum_{j ,l =0}^{N-1, M-1} f[j,l] e^{-i 2 \pi (\frac{ k j}{N} + \frac{ m l}{M})} \hspace{1cm}
	f[j, l] :=\frac{1}{NM} \sum_{k ,m =0}^{N-1, M-1} F[k,m] e^{i 2 \pi(\frac{ k j}{N} + \frac{m l}{M})} . \label{eq:2d-dft}
\end{align}
\end{problem}\label{prob:2DPR}
\noindent

Unlike the one-dimensional problem, phase retrieval in two or more dimensions almost always has a unique solution \cite{kogan}.
This is the motivation behind studying a $2D$ version of the problem, and in demonstrating its applicability to the problem of quantum phase estimation.
The uniqueness properties of a signal can be studied through setting up an associated polynomial, as discussed throughout the phase retrieval literature, \cite{kogan}.
Whether these polynomials are irreducible or not determines whether the problem has non-trivial ambiguities.
Almost all $2D$ multivariate polynomials are irreducible, a fact which means almost all $2D$ signals can be recovered from their Fourier magnitudes \cite{kogan}. 
This is our motivation for demonstrating one possible way of embedding quantum phase estimation into a $2$D phase retrieval problem.
We will demonstrate one possible version of this, though many others can be imagined.

\subsection{The Hybrid Input-Output Algorithm}
There are many algorithms which can be used to solve the problem of phase retrieval \cite{phaselift,recent}.
We leave the exploration of these approaches to future work and instead simply demonstrate the effectiveness using one of the basic algorithms in widespread use, Fienup's hybrid input-output algorithm or HIO algorithm \cite{fienup82}.

\begin{figure}[t]
	\begin{subfigure}{.45\linewidth}
		\centering
		\begin{algorithmic}[1]
			\Procedure {Hybrid}{$|f|$, $\beta$, $L$, $F^1$}
			
			\For {i = 1, 2, \ldots L}
			\State $f^i = \mathcal{DFT}^{-1} (F^i) = |f^i| e^{i \text{arg}(f^i)}$
			\State $\tilde{f}^i = |f| e^{i \text{arg}(f^i)}$
			\State $\tilde{F}^i = \text{real}(\mathcal{DFT} (\tilde{f}^i))$
			\If{$\tilde{F}^i[k,m] \leq 0$}
			\State $F^{i+1}[k,m]=F^{i}[k,m] - \beta \tilde{F}^{i}[k,m]$
			\Else
			\State $F^{i+1}[k,m]=\tilde{F}^i[k,m]$
			\EndIf
			\EndFor
			\EndProcedure
		\end{algorithmic}
	\end{subfigure}
	\hspace{1cm}
	\begin{subfigure}{.45\linewidth}
		\centering
		\begin{tikzpicture}[node distance=2.0cm, auto]
			\node (tl)[draw] {$F^{i}$};
			\node (t) [right of=tl,draw,fill=Apricot] {$\mathcal{DFT}^{-1}$} ;
			\node (tr) [right of=t,draw] {$f^{i}$} ;
			\node (cl) [below of=tl,fill=Apricot,draw] { Real and Positive};
			\node (bl) [below of=cl,draw] {$\tilde{F}^{i}$};
			\node (b) [right of =bl,fill=Apricot,draw]{$\mathcal{DFT}$};
			\node (br) [right of =b,draw]{$\tilde{f}^{i}$};
			\node (cr) [below of = tr,fill=Apricot,draw]{$|f[j,l]|$};
			\draw[->] (tl) to node {} (t);
			\draw[->] (t) to node {}(tr);
			\draw[->] (tr) to node {}(cr);
			\draw[->] (cr) to node {}(br);
			\draw[->] (br) to node {}(b);
			\draw[->] (b) to node {}(bl);
			\draw[->] (bl) to node {}(cl);
			\draw[->] (cl) to node {}(tl);
		\end{tikzpicture}
	\end{subfigure}
	
	\caption{a) Schematic for Fienup's hybrid input-output algorithm four steps. The $i-$th iteration starts with a candidate spectrum $F^i$: Step (1) transform it into a time series $f^i$ candidate via an inverse DFT; Step (2) generates a new $\tilde{f}^i$ that has same phase as $f^i$
    and satisfies $|\tilde{f}[j,l]|=|f[j,l]|$; Step (3) transform it back to the Fourier domain; Step (4) take the real part and then
    implement the update rule in the schematic to impose positivity of the spectrum.}\label{fig:hybrid-pscode}
\end{figure}
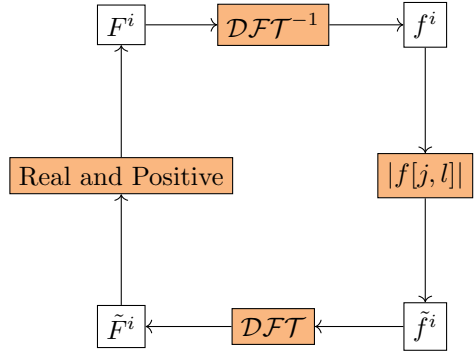

The HIO algorithm is a generalisation of the earlier Gerchberg-Saxton algorithm. 
Figure \ref{fig:hybrid-pscode}  summarizes the  algorithm which involves transforming back and forth between the Fourier and object domain and imposing constraints on each domain at every iteration. The signal constraints being the measured magnitudes, while the Fourier domain constraints are more flexible. Sometimes knowledge of the support of $F$ may be used, or knowledge that $F[k,m]$ is real and positive. We will use these latter two constraints from now on.

Let $F^i$ denote the estimate for the true DFT $F$ at step $i$ in the algorithm. 
An iteration of the algorithm can be summarised as follows.
\begin{enumerate}
    \item The inverse discrete Fourier transform of $F^i$ is taken to obtain $f^i = \mathcal{DFT}^{-1} (F^i) = |f^i| e^{i \text{arg}(f^i)}$. 
    \item The absolute values $|f^i|$ are replaced with the true and known absolute values $|f|$.
    This defines $\tilde{f}^i = |f| e^{i \text{arg}(f^i)}$, a function which satisfies the absolute values constraints.
    \item Next one takes the DFT of this time series, and enforces the constraint that the DFT is \textbf{real}, to obtain $\tilde{F}^i = \text{real}(\mathcal{DFT} (\tilde{f}^i))$.
    \item Next the \textbf{positivity} constraint is used. One defines each element of $F^{i+1}$ with the following update rule,
\begin{align}
	F^{i+1}[k,m]
	&= \begin{cases}
		\tilde{F}^{i}[k,m] ,& \tilde{F}^{i}[k,m] \geq0\\
		F^{i}[k,m] - \beta \tilde{F}^{i}[k,m],  & \text{else}
	\end{cases}.
\end{align}
This retains the elements which are positive and replaces the remaining entries by a relaxed version of the positivity constraint with parameter $0\leq\beta\leq 1$. 

\end{enumerate}
 The algorithm then loops for $L$ iterations, hopefully converging to the true $F[k]$ or a good approximation of it.

\subsection{Adapting 2D Quantum Signals for the HIO Algorithm}
\subsubsection{Extension to 2D}
In this section we first explain how to embed spectral quantum phase-estimation, which is a $1$D problem, into a $2$D signal processing problem and then show that we can add constraints to this signal to ensure that the target $F$ for the HIO algorithm is real and positive.
For the purposes of quantum phase estimation we are interested in $1$D signals of the form $f(t) = \bra{\psi}e^{i t H} \ket{\psi}$, where $H$ is the Hamiltonian being studied.
To create a $2D$ problem we introduce a secondary time-evolution over a dummy Hamiltonian $H_D$ with virtual time parameter $z$, defining the new continuous $2D$ signal:
\begin{align}\label{eq:def-F-cont}
	f(t, z) \coloneqq \bra{\psi} e^{i t H }e^{iz H_D} \ket{\psi}.  
\end{align}
It is important to choose $H_D$ such that $\ket{\psi}$ is not an eigenstate of it, or otherwise  
the problem remain of a $1$D nature.
This is one of the simplest ways we can imagine creating a $2$D signal.

\subsubsection{Continuous to Discrete}
First we adapt some key concepts in Section \ref{sec:prelims}
on continuous-time signals, their Fourier transforms and their discretizations to the two-dimensional setting.
We establish these facts only so that we can show that the discrete Fourier transform $F$ can be guaranteed to be real and positive.
This is key to ensuring the reliability of the HIO algorithm. 

Now the time signal windowed and sampled time-series $g(t,z)$ and its Fourier transform $G(\omega, \eta)$ now read
\begin{align}\label{eq:2d-contin}
    g(t,z) &:= f(t,z) \Pi_T(t,z) \Sh_{\Delta t} (t,z) \\
    G(\omega,\eta) &:= \mathcal{FT}[g(t,z)](\omega,\eta).
\end{align}
The two dimensional Dirac comb with $\Delta t = T/N$ which represents sampling in the continuous picture, and two dimensional windowing function are
\begin{align}
    \Sh_{\Delta t}(t, z) &= \sum_{j,l = -\infty}^{\infty} \delta(t - j \Delta t ) \delta(z - l \Delta t ) \\
    \Pi_T(t,z) &= \Lambda_{T}(t) \cdot \Lambda_{T} (z ) 
	%\Lambda_{T}(t) &= \begin{cases}
	%	1 - 2| \frac{t}{ T}| & |t| \leq  T/2 \\
		%0 & \text{else}
	%\end{cases}
\end{align}
where $\Lambda_{T}(t)$ and $\Lambda_{T}(z)$ have support for times  $t, z \in [-T/2, T/2]$ as before. 
In the above review and for the remainder of this analysis we take $M=N$ for simplicity, though what follows is readily generalized to $M \neq N$.

As established in Section \ref{sec:prelims}, the discrete and finite matrix $F[k,m]$ is related to the continuous time Fourier transform $G(\omega, \eta) = \mathcal{FT}[g(t,z)](\omega,\eta)$ at the discrete frequencies $\omega_k = 2 \pi k/T$ and $\eta_m = 2 \pi m/T$ via 
\begin{align}\label{eq:F-connect}
    F[k,m] &:= G(\omega_k,\eta_m) ,
\end{align}
with $k = 0, 1, \ldots N-1$ and $m = 0 , 1, \ldots M-1$.
Similarly $f[j,l]$ is related to this underlying continuous picture via 
\begin{align}\label{eq:f-connect}
    \hat{f}[j,l]&:= f(j \Delta t, l \Delta  t) \Pi_T(j \Delta t, l \Delta t) \\
    f[j,l] &:= \sum_{j',l' \in \mathbb{Z}} \hat{f}[j - j' N,l-l'M] 
\end{align}
where the $\hat{f}[j,l]$ is defined for all $j,l \in \mathbb{Z}$ and $f[j,l]$ is defined for $j = 0, 1, \ldots N-1$ and $l = 0 , 1, \ldots M-1$. 
In summary, we have two underlying continuous functions related by a continuous two-dimensional Fourier transform $G(\omega,\eta) = \mathcal{FT}[g(t,z)](\omega, \eta)$ and two matrices related by a two-dimensional discrete time Fourier transform $F[k,m] = \mathcal{DFT}[f[j,l]]$ and we can relate these two pictures via Eq.(\ref{eq:F-connect}) and Eq. (\ref{eq:f-connect}).

\subsubsection{Target DFT is Real and Positive}
Now we can use the relationship between the continuous and discrete pictures to show that we can define a $2$D phase retrieval problem where the DFT we wish to reconstruct is both real and positive. 
Knowing this, justifies using these facts as constraints in the HIO algorithm to get reliable performance.

\textbf{Positivity Constraint:} First we can show that $F[k,m]$ will be positive when we choose $H_D$ to commute with $H$ and as a consequence of choosing a triangular windowing function
\begin{align}
    \Pi_T(t,z) &= \Lambda_{T}(t) \cdot \Lambda_{T} (z ) \\
	\Lambda_{T}(t) &= \begin{cases}
		1 - 2| \frac{t}{ T}| & |t| \leq  T/2 \\
		0 & \text{else}.
	\end{cases}
\end{align}
This follows from establishing that $G(\omega,\eta)$ is positive, which then ensures $F[k,m]$ is positive  as it is related to this continuous function via Eq. \ref{eq:F-connect}.
One can write $G(\omega,\eta)$ as a convolution as
\begin{align}\label{eq:G-conv}
    G(\omega,\eta) &= \mathcal{FT}[f(t,z) ] \ast \mathcal{FT}[\Pi_T(t,z)] \ast \mathcal{FT}[ \Sh_{\Delta t} (t,z)].
\end{align}
We will evaluate each term in this convolution in turn.
First we have
\begin{align}
    \mathcal{FT}[f(t,z) ] &= \int_{-\infty}^{\infty} \int_{-\infty}^{\infty}f(t,z) e^{- i \omega t} e^{- i \eta z} dt dz \\
    &=\int_{-\infty}^{\infty}\int_{-\infty}^{\infty} \bra{\psi} e^{i t H + i z H_D} \ket{\psi} e^{- i \omega t} e^{- i \eta z} dt dz.
\end{align}
Now we use the fact that $[H,H_D] = 0$ to write both Hamiltonians in their shared eigenbasis as $H =\sum_i E_i \Pi_i$ and $H_D =\sum_i E^D_i \Pi_i$, where $\Pi_i = \ket{E_i} \bra{E_i}$ is the projector onto this share eigenbasis. 
This gives
\begin{align}
    \mathcal{FT}[f(t,z) ] &=\int_{-\infty}^{\infty}\int_{-\infty}^{\infty} \bra{\psi}\Pi_i\ket{\psi} e^{i t E_i + i z E^D_i}  e^{- i \omega t} e^{- i \eta z} dt dz \\
    &=\sum_i |\braket{\psi |E_i}|^2  \int_{-\infty}^{\infty}  e^{i t (E_i - \omega)}dt  \int_{-\infty}^{\infty}e^{ i z (E^D_i- \eta )} dz \\
    &=4 \pi^2\sum_i |\braket{\psi |E_i}|^2 \delta(\omega - E_i) \delta(\eta - E_i^D).
\end{align}
This function is positive for all $\omega$ and $\eta$. This is why we have introduced a commuting dummy Hamiltonian. 
If $H$ and $H_D$ do not commute is is possible to have negative values in this step. 

The second term to evaluate is the Fourier transform of the windowing function.
This is
\begin{align}
    \mathcal{FT}[\Pi_T(t,z) ] &=\int_{-\infty}^{\infty}\int_{-\infty}^{\infty} \Lambda_T(t) \Lambda_T(z)  e^{- i \omega t} e^{- i \eta z} dt dz \\
    &=\frac{8 \sin ^2\left(\frac{T \omega }{4}\right)}{T \omega ^2} \frac{8 \sin ^2\left(\frac{T \eta }{4}\right)}{T \eta ^2},
\end{align}
which we can see is positive for all $\omega$ and $\eta$.  Note that one could have used any windowing function with a positive Fourier transform and better choices than a triangular windows deserve further investigation.
%We have chosen a simple triangular window as its Fourier transform is positive as in shown here. 
Finally, as discussed in Section \ref{sec:prelims} (the Fourier transform of a Dirac comb is another Dirac comb), the effect of $\mathcal{FT}[ \Sh_{\Delta t} (t,z)]$ on eq. (\ref{eq:G-conv}),
we can establish that $G(\omega, \eta)$ is a positive function and therefore that $F[k,m]=G(\omega_k, \eta_m)$ is always positive. This allows us to use this as a constraint to drive the convergence of the HIO algorithm.

\textbf{Real Constraint:} 
We can show that $F$ is real very simply, thus making this a second suitable driving constraint.
In order for $F$ to be real the $2$D signal must satisfy 
\begin{align}
    f[j,l] = f^*[N-j,M-l], \label{eq:real-cond}
\end{align}
modulo $N$ and $M$ and where $j = 0,1, \ldots N-1$ and $l=0,1, \ldots M-1$.
This is a generic fact about the DFT and can be found in any textbook and easily verified. Any signal which satisfies this condition will have a real discrete Fourier transform.

We can now show that our signal satisfies this condition using the fact that
\begin{align}
    \hat{f}[j,l] = \hat{f}^*[-j,-l],
\end{align}
which follows from the definition of $\hat{f}[j,l] = f(t_j, z_l) \Pi_T(t_j, z_l)$ and the fact that the continuous functions satisfy $f(t, z)=f^*(-t, -z)$ and $\Pi_T(t, z)=\Pi_T^*(-t, -z)$.

We can combine this with the definition of $f[j,l]$ to show that it satisfies Eq. \ref{eq:real-cond}. Writing this out explicitly we see that
\begin{align}
    f^*[N-j,M-l] &= \sum_{l',j' \in \mathbb{Z}} \hat{f}^*[N-j - Nj',M-l - Ml'] \\
    &=\sum_{l'',j'' \in \mathbb{Z}} \hat{f}^*[-(j - Nj''),-(l - Ml'')] \\
    &=\sum_{l'',j'' \in \mathbb{Z}} \hat{f}[j - Nj'',l - Ml'']\\
    &= f[j,l].
\end{align}
Therefore $f[j,l]$ satisfies Eq. \ref{eq:real-cond} and therefore  its discrete Fourier transform $F[k,m]$ is real. Therefore we can use this fact to drive the HIO algorithm.

\subsection{Implementation and Numerical Results}

We test the performance of this algorithm for instances of the Fermi-Hubbard model as explained in Section \ref{sec:prelims}. We take as target input state $\ket{\psi}$ the uniform superposition over computational basis states. As dummy Hamiltonian $H_D$ we choose the total number operator 
\begin{align}
    H_D = \sum_{i \in V, \sigma}a^\dag_{i\sigma} a_{i\sigma}.
\end{align}
Under the Jordan-Wigner encoding this maps to
\begin{align}
    H_D \rightarrow \frac{1}{2} \sum_{i \in V, \sigma}\left(I_{i\sigma} - Z_{i,\sigma}\right).
\end{align}
This makes $e^{i t H_D}$ trivial to implement via a single layer of single qubit rotation gates. 
When $\ket{\psi}$ is a superposition of computational basis states we then  
satisfy the requirement that $\ket{\psi}$ must not be eigenstate of $H_D$.

\subsubsection{Image Reconstruction}
First we demonstrate that the HIO algorithm works well given quantum signals set up according to the previous section.
We test this numerically using a $2$D spin Fermi-Hubbard model as $H$ in Eq. (\ref{eq:def-F-cont}), with the hopping strength $\tau=1$ and on-site interactions $U=4$.

\begin{figure}[h]
	\centering
	\centering
\begin{subfigure}{0.329\textwidth}
    \includegraphics[width=\textwidth]{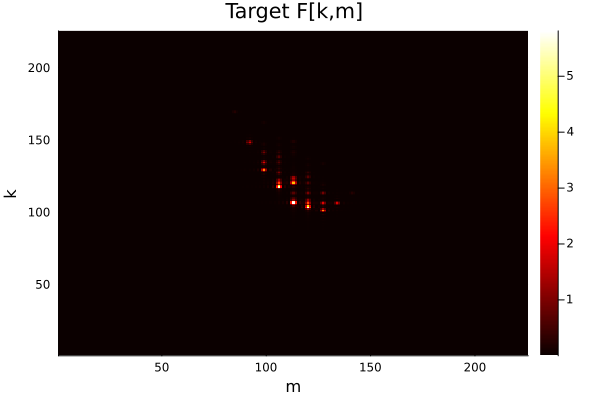}
    \caption{Target spectrum.}
    \label{fig:first}
\end{subfigure}
\hfill
\begin{subfigure}{0.329\textwidth}
    \includegraphics[width=\textwidth]{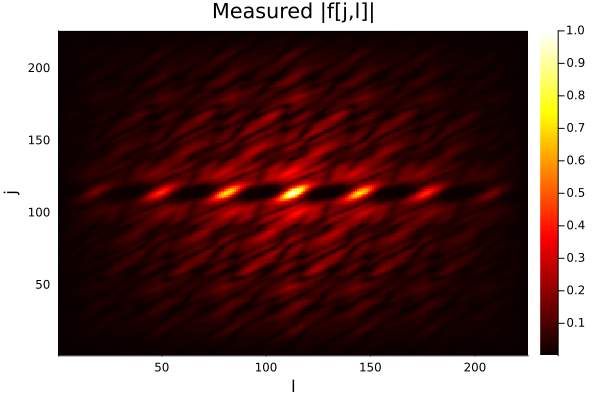}
    \caption{Measured time-series.}
    \label{fig:second}
\end{subfigure}
\hfill
\begin{subfigure}{0.329\textwidth}
    \includegraphics[width=\textwidth]{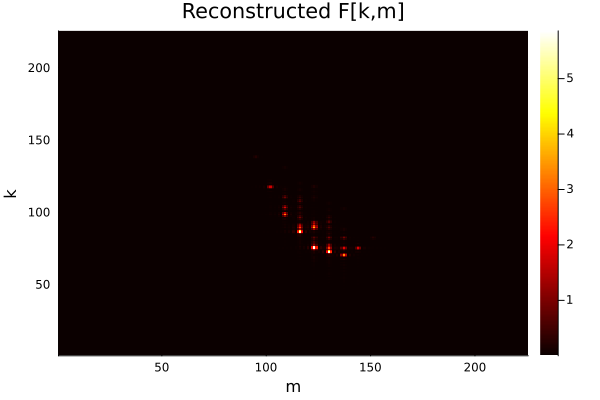}
    \caption{Reconstructed spectrum.}
    \label{fig:third}
\end{subfigure}
	\caption{
		An example of $2$D phase retrieval for a $2\times2$ Fermi-Hubbard Hamiltonian with $N \times M = \N \times \M$. We take $\ket{\psi}$ to be a uniform superposition over computational basis states. The true signal $f[j,l]$ is shown left, the absolute values $|F[k,m]|$ are shown in the centre and finally the reconstructed $f[j,l]$ is shown to the right, where it has been recovered up to a trivial ambiguity corresponding to a translation.
	}\label{fig:2d-22}
\end{figure}

In addition to the constraints that the signal is real and positive, we also incorporate knowledge of the phases for $f[0,l]$.
This information can be computed classically in any given experiment as these points correspond to evolution under $H_D$ only, provided the initial state it acts upon is also easy to classically simulate.

Here we show a typical example of signal reconstruction in Figure \ref{fig:2d-22} for a $2 \times 2$ Fermi-Hubbard model.
For this case we have set $N = M = \N$, time $T  =\T$ and chosen $\ket{\psi}$ as the uniform superposition over computational basis states. $H$ has been normalized so that its eigenvalues lie between $0$ and $\pi$.

The left of Fig. \ref{fig:2d-22} displays in a colourmap the value of the ideal two-dimensional time-series $F[k,m]$ obtained numerically through brute force simulation. In the centre of Fig. \ref{fig:2d-22} we show the absolute values $|f[j,l]|$. This is the input to the HIO algorithm and is what would be measured via a quantum circuit implementing time evolution on the state $\ket{\psi}$.
Observe that we also see the effect of the $2D$ triangular windowing function, so that the magnitude of $|f[j,l]|$ decreases towards the outer edges of the plot.
Finally the reconstructed signal is show on the right of Fig. \ref{fig:2d-22}.
This was obtained with $\beta=\bval$ and $L=\iter$ iterations of the HIO algorithm.
The reconstruction is qualitatively good and in agreement with the true signal for the noiseless case, up to a translation, consistent with the trivial ambiguities discussed above and that can be compensated for.  

\subsubsection{Spectrum recovery}
\begin{figure}[h]
	\centering
	\begin{subfigure}{0.45\linewidth}
		\centering
		\includegraphics[width=\linewidth]{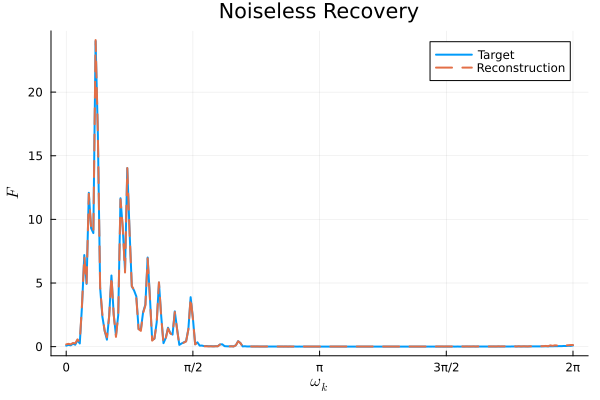}
	\end{subfigure}
	\begin{subfigure}{0.45\linewidth}
		\centering
		\includegraphics[width=\linewidth]{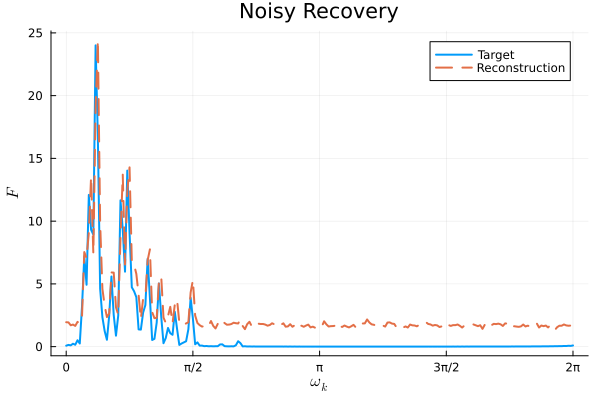}
	\end{subfigure}
	
	\caption{The results of plotting $F[k]:=\mathcal{DFT}[f[j,0]]$.  Here we show the result for the equal superposition over all computational basis states, with $T  =\T$, $N=\N$ and $M=\M$ and $H$ has been normalized so that its eigenvalues lie between $0$ and $\pi$.
    Here the ideal solution includes the triangular window by definition.
    On the left we show the noiseless case and on the right we show the case for $1000$ samples per time-series point. The true and recovered signals are shown and overlap almost exactly in the noiseless case.
	}\label{fig:spectra}
\end{figure}

As mentioned earlier, the last step is to extract the desired information about the spectrum of $H$ from the recovered signal by computing the DFT of the first column of the two dimensional time-series matrix, i.e., $F[k]:=\mathcal{DFT}[f[j,0]]$, corresponding to a zero virtual time evolution $z=0$.
The result of this is shown in Fig. \ref{fig:spectra}.
On the left we reconstruct the spectrum from the ideal noiseless time-series, were in the right we  included the effect of sampling noise, where we model a shot noise corresponding to $1000$ samples (circuit shots) per required expectation value.

We compare the phase-retrieval recovery (orange) to the reference signal (blue) obtained with an ideal Hadamard-like quantum circuit and using statistical QPE with same parameters $T$ and $\Delta t$,
including the effect of the triangular window.

 To obtain these plots we have shifted the recovered signal so that the trivial ambiguity is eliminated.
This is to demonstrate the degree of overlap between the true and recovered signal up to this ambiguity.
We see that in the noiseless case the spectrum is recovered well.
When we introduce sampling noise the shape and location of the peaks in the spectrum are recovered, but there is a constant positive offset in the spectrum, that could potentially be filtered out without damaging the reconstructed spectrum too much.

\section{Appendix D: Comparison with the standard approach}
\label{sec:comparison}

There are two key advantages of phase retrieval compared with the standard approach to statistical phase estimation: each qubit no longer needs to interact with an ancilla ``control'' qubit, and there is no increase in circuit complexity caused by the use of controlled unitary operations. One can consider many metrics for circuit complexity; here, we will focus on the number of CNOT gates required, and the quantum circuit depth.

To get a sense of how large the reduction in complexity can be, first observe that if the quantum hardware allows for all-to-all connectivity, any quantum circuit computing a unitary operator $U$ can be converted into a quantum circuit for computing controlled-$U$ by replacing every CNOT gate with a Toffoli gate, and every single-qubit unitary $V$ with a controlled-$V$ operator. A Toffoli gate can be implemented using 6 CNOT gates and controlled-$V$ can be implemented using 2 CNOT gates for any single-qubit unitary $V$. So, if the original quantum circuit had $c$ CNOT gates and $s$ single-qubit gates, the controlled circuit will have at most $6c + 2s$ CNOT gates -- at most a constant factor increase in complexity, if $c \ge s$ (for example), but still relatively substantial for near-term applications. However, the difference in complexity between the two situations can be greater if there are restrictions on hardware connectivity, or if one considers circuit depth instead.

Here we approximately calculate the reduction in circuit complexity achieved by phase retrieval for several simple examples: the 1D Ising model with transverse field, both all-to-all and with matching hardware connectivity, and the spinless 2D Fermi-Hubbard model with square-lattice hardware connectivity. In each case, we consider the complexity of implementing $k$ Trotter steps -- i.e.\ $k$ repetitions of an operator of the form $\prod_{j=1}^m e^{i \theta_j H_j}$ for a Hamiltonian of the form $H = \sum_j H_j$. Note that it is important to consider $k > 1$ steps because the cost of interacting with the control qubit can be amortised across multiple steps by preparing a GHZ state, as we will see below. This approach may add its own difficulties in terms of rendering the circuit more prone to decoherence, but we will ignore this issue for simplicity.
We also stress that there is no guarantee that the implementations we describe here are optimal.

\begin{table}[h]
\begin{center}
    \begin{tabular}{|c|c|c|c|c|c|}
        \hline Model & H/w & CNOTs (PR) & CNOTs (no PR) & Depth (PR) & Depth (no PR) \\
        \hline 1D TFIM & All-to-all & $(2n-2)k$ & $(6n-4)k$ & $4k$ & $2\lceil \log_2 n \rceil + 10k$\\
        1D TFIM & 1D & $(2n-2)k$ & $6(n-1) + (6n-4)k$ & $4k$ & $6\lceil n/2 \rceil + 10k$\\
      %  2d TFIM & 2d & & & &\\
        2D FH & 2D & $32 k n (n-1)/2  $ & $48 k n (n-1)/2   +6 \lceil (n-1)^2 / 2 \rceil$& $32 k$ & $48 k+3 (n-2)$\\
        \hline 
    \end{tabular}
\end{center}
\caption{The cost of implementing $k$ Trotter layers for several Hamiltonians, an $n$ qubit transverse field Ising model and an $n\times n$ spinless Fermi-Hubbard model. ``Depth'' is CNOT depth. Note that the algorithms achieving the minimal CNOT depth and CNOT count may be different. We consider the spinless Fermi-Hubbard model for simplicity, and these calculations are approximate.}
\label{tab:complexities}
\end{table}
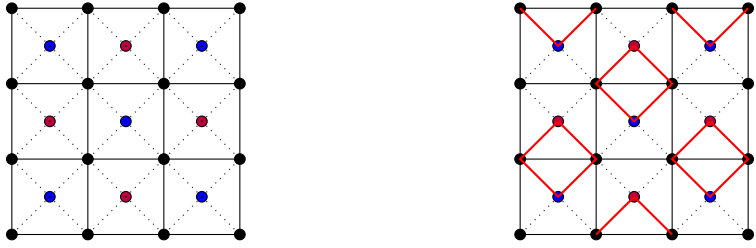
\begin{figure}
 \centering
\begin{subfigure}{0.4\textwidth}
    \centering
    	\begin{tikzpicture}
		
		% Draw the vertical lines
		\foreach \x in {0,1,2,3} {
			\draw (\x,0) -- (\x,3);
		}
		
		% Draw the horizontal lines
		\foreach \y in {0,1,2,3} {
			\draw (0,\y) -- (3,\y);
		}
		
		% Draw points at the lattice intersections (black nodes)
		\foreach \x in {0,1,2,3} {
			\foreach \y in {0,1,2,3} {
				\filldraw[fill=black] (\x,\y) circle (2pt); % Black nodes at intersections
			}
		}
		
		% Draw nodes at the center of each square (faces)
		\foreach \y in {0.5, 1.5, 2.5} {
			\foreach \x in {0.5, 1.5, 2.5} {
				% Switch color order based on row (y-coordinate)
				\ifodd\y
				% Row 1 (y=0.5) is blue and Row 3 (y=2.5) is blue
				\ifodd\x
				\filldraw[fill=blue] (\x,\y) circle (2pt); % Blue for odd x in odd rows
				\else
				\filldraw[fill=purple] (\x,\y) circle (2pt); % Purple for even x in odd rows
				\fi
				\else
				% Row 2 (y=1.5) is purple and Row 3 (y=2.5) is blue
				\ifodd\x
				\filldraw[fill=purple] (\x,\y) circle (2pt); % Purple for odd x in even rows
				\else
				\filldraw[fill=blue] (\x,\y) circle (2pt); % Blue for even x in even rows
				\fi
				\fi
			}
		}
		
		% Connect face nodes to black nodes with dotted lines
		\foreach \x in {0.5, 1.5, 2.5} {
			\foreach \y in {0.5, 1.5, 2.5} {
				% Connect to the corresponding black nodes
				\draw[dotted] (\x,\y) -- (\x-0.5,\y-0.5); % Connect to bottom-left black node
				\draw[dotted] (\x,\y) -- (\x-0.5,\y+0.5); % Connect to top-left black node
				\draw[dotted] (\x,\y) -- (\x+0.5,\y-0.5); % Connect to bottom-right black node
				\draw[dotted] (\x,\y) -- (\x+0.5,\y+0.5); % Connect to top-right black node
			}
		}
		
	\end{tikzpicture}
 \end{subfigure}
\centering
\begin{subfigure}{0.4\textwidth}
    \centering
    	    	\begin{tikzpicture}
		
		% Draw the vertical lines
		\foreach \x in {0,1,2,3} {
			\draw (\x,0) -- (\x,3);
		}
		
		% Draw the horizontal lines
		\foreach \y in {0,1,2,3} {
			\draw (0,\y) -- (3,\y);
		}
		
		% Draw points at the lattice intersections (black nodes)
		\foreach \x in {0,1,2,3} {
			\foreach \y in {0,1,2,3} {
				\filldraw[fill=black] (\x,\y) circle (2pt); % Black nodes at intersections
			}
		}
		
		% Draw nodes at the center of each square (faces)
		\foreach \y in {0.5, 1.5, 2.5} {
			\foreach \x in {0.5, 1.5, 2.5} {
				% Switch color order based on row (y-coordinate)
				\ifodd\y
				% Row 1 (y=0.5) is blue and Row 3 (y=2.5) is blue
				\ifodd\x
				\filldraw[fill=blue] (\x,\y) circle (2pt); % Blue for odd x in odd rows
				\else
				\filldraw[fill=purple] (\x,\y) circle (2pt); % Purple for even x in odd rows
				\fi
				\else
				% Row 2 (y=1.5) is purple and Row 3 (y=2.5) is blue
				\ifodd\x
				\filldraw[fill=purple] (\x,\y) circle (2pt); % Purple for odd x in even rows
				\else
				\filldraw[fill=blue] (\x,\y) circle (2pt); % Blue for even x in even rows
				\fi
				\fi
			}
		}
		
		% Connect face nodes to black nodes with dotted lines
		\foreach \x in {0.5, 1.5, 2.5} {
			\foreach \y in {0.5, 1.5, 2.5} {
				% Connect to the corresponding black nodes
				\draw[dotted] (\x,\y) -- (\x-0.5,\y-0.5); % Connect to bottom-left black node
				\draw[dotted] (\x,\y) -- (\x-0.5,\y+0.5); % Connect to top-left black node
				\draw[dotted] (\x,\y) -- (\x+0.5,\y-0.5); % Connect to bottom-right black node
				\draw[dotted] (\x,\y) -- (\x+0.5,\y+0.5); % Connect to top-right black node
			}
		}
		\draw[thick, red] (0.5, 0.5) -- (1, 1) -- (0.5, 1.5) -- (0, 1) -- cycle;
        \draw[thick, red] (1.5, 1.5) -- (2, 2) -- (1.5, 2.5) -- (1, 2) -- cycle;
        \draw[thick, red] (2.5, 0.5) -- (3, 1) -- (2.5, 1.5) -- (2, 1) -- cycle;
        \draw[thick, red]    (0, 3) -- (0.5, 2.5) -- (1, 3) ;
        \draw[thick, red]    (2, 3) -- (2.5, 2.5) -- (3, 3) ;
        \draw[thick, red]    (1, 0) -- (1.5, 0.5) -- (2, 0) ;
	
	\end{tikzpicture}
 \end{subfigure}

    \caption{A possible layout of the spinless Fermi-Hubbard model on a lattice with $2$D connectivity. The connectivity is indicated by dashed lines while the Fermi-Hubbard lattice interactions are indicated by solid lines. The face qubits in purple are needed for the compact encoding. The face qubits in blue are used to implement geometrically local control. The diagram on the right indicates one of the four Trotter layer groupings of interactions which can be implemented in parallel, in this case horizontal hopping interactions. With the control and encoding qubits these interactions are at most Pauli weight-$4$. }
    \label{fig:FH-layout}
\end{figure}
\begin{enumerate}
    \item \textbf{1D Ising model with transverse field.} This Hamiltonian is defined as $H = \sum_{i=1}^{n-1} Z_i Z_{i+1} + \sum_{j=1}^n X_j$. A Trotter step can be implemented using $2(n-1)$ CNOT gates and in CNOT depth $4$. To implement a controlled Trotter step, we need to implement controlled-$e^{i\theta ZZ}$ and controlled-$e^{i\theta X}$ gates. These have circuits using four and two CNOT gates, respectively. With all-to-all connectivity, then, each Trotter layer can be implemented using $4(n-1)+2n = 6n-4$ CNOT gates. Implementing these in a straightforward way would lead to a high-depth circuit, as each controlled unitary acts on the same control qubit. However, we can reduce the depth by constructing a GHZ state of $n$ qubits, where each gate that needs to be applied in parallel uses a different qubit of this state. The CNOT depth of producing such a state given all-to-all connectivity is at most $\lceil \log_2 n \rceil$. Following this, we can split the time-evolution steps into three groups (ZZ terms on (odd, even) qubits, ZZ terms on (even, odd) qubits, and X terms) and implement each group in parallel. At the end, the GHZ state is uncomputed, before measuring the control qubit as in the usual protocol. The total CNOT depth (with all-to-all connectivity) is then $2\lceil \log_2 n \rceil + 10k$ as each layer can be implemented in CNOT depth $10$.

    With $1$D connectivity, the situation is more difficult as we need to create a distributed GHZ state. This can be done with a small extra cost in gate count, but a linear cost in depth. Now we include an ancilla qubit for controlled operations next to each qubit in the original $1$D graph. Once these qubits have been prepared in a GHZ state, we can perform controlled operations across any pair of qubits in the original graph without the need for any swaps. The map $\ket{000} \mapsto \ket{000}$, $\ket{100} \mapsto \ket{101}$ can be carried out using three CNOT gates in 1d connectivity (CNOT$_{12}$, CNOT$_{23}$, CNOT$_{12}$), so the required state can be prepared using $3(n-1)$ CNOT gates and in CNOT depth $3\lceil n/2 \rceil$. Once this has been prepared, the costs for the time-evolution part are the same as for all-to-all connectivity. The total CNOT cost for $k$ Trotter steps, including the cost of uncomputing the GHZ state, is then $6(n-1) + (6n-4)k$, and the CNOT depth is $6\lceil n/2 \rceil + 10k$.

    \item \textbf{2D Fermi-Hubbard model.} Consider an $n\times n$ Fermi-Hubbard model without spin on an $2$D lattice. Consider even $n$ only and use the compact fermionic encoding~\cite{Derby21} and layout shown in Figure \ref{fig:FH-layout}. The full Hamiltonian is made of horizontal $\sum_{\langle i,j \rangle} (X_iX_jY_{f(i,j)} + Y_iY_jY_{f(i,j)})/2$ and vertical $\sum_{\langle i,j \rangle} \pm(X_iX_jX_{f(i,j)} + Y_iY_jX_{f(i,j)})/2$ interactions, where the sum is over pairs of horizontally or vertically adjacent sites on an $n\times n$ lattice and $f(i,j)$ are face qubits placed on this lattice. The Hamiltonian can be split into four sets of $n (n-1)/2$ interactions. Within each group all interactions can be performed in parallel. Every interaction consists of two weight-$3$ (at most) Pauli interactions. Therefore each interaction can be done using $8$ CNOT gates, meaning that a full Trotter step will have a depth of $32$ CNOT gates and the total number of CNOT gates will be $32 n (n-1)/2$. This calculation is approximate and simply treats edge interactions identically to those in the bulk.
    When performing controlled evolution under this Hamiltonian one possible layout of control ancilla qubits is shown in Figure \ref{fig:FH-layout}. As in the case of the Ising model above, we consider creating a distributed GHZ state so that we can implement local control. This particular selection uses the remaining face qubits not used by the compact encoding. Now every "controlled" interaction can be performed by evolving under two weight-$4$ Pauli terms, where the interactions have been augmented so that they include a Pauli-$Z$ on the local control ancilla. Therefore each interaction can be done using $12$ CNOT gates, meaning that a full Trotter step will have a depth of $48$ CNOT gates and the total number of CNOT gates will be $48 n (n-1)/2 $. (This all uses the standard approach to performing a controlled rotation by a weight-k Pauli operation \cite{NC}) The additional cost of creating the distributed GHZ state depends on the number of control ancillas $\lceil (n-1)^2 / 2 \rceil$ and their connectivity. The total count of CNOT gates will be $3 \times \lceil (n-1)^2 / 2 \rceil$. The depth is approximately $3 (n-2)/2$ (for even lattices with $n>2$). One can begin building the GHZ state from the central control qubit and move outward along the branches approximately in parallel.
    It takes $3$ CNOT gates to spread the entanglement to the nearest control qubit, shown in blue. This is done in the same manner as described for the Ising model.
   
\end{enumerate}

In all of the above costings, we do not include the cost of preparing (or uncomputing) the initial state $\ket{\psi}$ -- which we assume to be simple -- nor the cost of preparing a superposition of $\ket{\psi}$ with another state. In some situations, this may be fairly substantial, e.g.\ if this superposition is a GHZ state.

\bibliography{refs}

\end{document}